\documentclass[%
 reprint,
 amsmath,amssymb,
 aps,
]{revtex4-2}

\newcommand{\angstrom}{\mbox{\normalfont\AA}}
\usepackage{graphicx}
\usepackage{dcolumn}
\usepackage{subcaption}
\usepackage{hyperref}

\usepackage{bm}
\usepackage[utf8]{inputenc}
\usepackage[T1]{fontenc}
\usepackage[dvipsnames]{xcolor}

\DeclareUnicodeCharacter{0301}{\'{e}}
\DeclareUnicodeCharacter{00E9}{\'{e}}
\DeclareUnicodeCharacter{00FC}{\"{u}}
\usepackage{xcolor}

\usepackage[nameinlink,capitalise]{cleveref}
\newcommand{\lammp}{{\sc lammp}}
\newcommand{\python}{{\sc python}}
\newcommand{\refprop}{{\sc refprop}}
\newcommand{\nest}{{\sc nest}}

\begin{document}

\preprint{APS/123-QED}

\title{Model for bubble nucleation efficiency of low-energy nuclear recoils in bubble chambers for dark matter detection }

\author{Xiang Li }
    \email{xli11@ualberta.ca}
\author{Marie-Cécile Piro}
  \email{mariecci@ualberta.ca}
\affiliation{%
 Department of Physics, University of Alberta, Edmonton, T6G 2E1, Canada
}%


\date{\today}

\begin{abstract}
Bubble chambers are promising technologies for detecting low-energy nuclear recoils from the elastic scattering of dark matter particle candidates. Bubble nucleation occurs when the energy deposition exceeds a specific threshold defined traditionally by the “heat-spike” Seitz threshold. In this paper, we report on a physical model that can account for observed discrepancies between the current Seitz model and the measured nucleation efficiency of low-energy nuclear recoils, which is necessary for interpreting dark matter signals. In our work, we combine molecular dynamics and Monte Carlo simulations together with the Lindhard model to predict bubble nucleation efficiency and energy thresholds for C$_3$F$_8$, CF$_3$I, and xenon with enhanced accuracy over the Seitz model when compared to existing experimental data. We use our model to determine the effect on cross-section limits for spin-dependent and spin-independent interactions and compare it to the current PICO dark matter experiment. Our technique can also be applied to estimate the efficiency of future target fluids where no experimental data are available. As an example, we predict the nucleation efficiency, the energy threshold and the cross-section limits in the spin-independent channel for the Scintillating Bubble Chamber experiment filled with superheated liquid argon.

\end{abstract}

\maketitle


\section{\label{sec:intro} Introduction}

Superheated liquids in bubble chambers have been established as an attractive target in experiments searching for nuclear recoils induced by weakly interacting massive dark matter particles (WIMPs) \cite{ Feng:2010gw}. Over the years, such experiments \cite{ COUPP:2012jrk, PICO:2015yox, Behnke:2016lsk, PICO:2017tgi, PICO:2019vsc} have set the most stringent cross-section limits on direct detection of WIMPs in dark matter searches. To date, the PICO experiment has set the strongest constraint on the WIMP-proton spin-dependent cross section \cite{PICO:2019vsc}. Existing and planned experiments employ bubble chambers filled with fluorocarbons \cite{PICO:2019vsc} and noble liquids such as xenon and argon \cite{Baxter:2017ozv, Giampa:2021wte, SBC:2021yal}.

Bubble chambers consist of a vessel, in which the pressure and temperature of the target fluid set the energy threshold to detect nuclear recoils due to WIMPs scattering off atomic nuclei. Bubble nucleation is induced by particles interacting with energy above the energy threshold, typically described by Seitz’s “heat-spike” model \cite{ Seitz:1958nva},  producing a small volume of liquid to rapidly transition into a vapor state that eventually grows into a macroscopic bubble. However, it is well known from past and current experiments that the true threshold for nucleation deviates from the predicted Seitz threshold \cite{PICO:2015yox}. Furthermore, in this idealized model, the nucleation efficiency of nuclear recoils is assumed to be a step function at the Seitz threshold in contradiction with experiments. Until these effects are fully understood, nucleation efficiencies and energy thresholds must be determined experimentally by performing dedicated neutron calibrations since they are crucial parameters for interpreting results from direct searches for WIMPs \cite{PICO:2022nyi}.

In our work, we aim to understand and explain the processes that contribute to the resolution function that shifts and convolutes the Seitz step threshold. We propose a physical model using molecular dynamics (MD) simulations to investigate the bubble nucleation threshold (Sec. \ref{sec:MD}), combined with a Monte Carlo simulation with stopping and range of ions in matter (SRIM), which includes the Lindhard factor correction (Sec. \ref{sec:MCLind}). This paper presents our results for the nucleation efficiency obtained for C$_3$F$_8$,  CF$_3$I, xenon and argon at different energy thresholds set by experiments (Sec. \ref{sec:NuEff}). Finally, we use our model to obtain the cross-section limits in the spin-dependent and spin-independent sectors (Sec. \ref{sec:CS_DM}) and compare it to the prediction of the Seitz model and the efficiency used by the PICO dark matter experiment. We also predict cross-section limits for the future Scintillating Bubble Chamber (SBC) experiment filled with superheated liquid argon.

\section{\label{sec:seitz}Seitz Threshold}

The energy required to break the metastable state of a superheated liquid comes from the microscopic energy deposition of incoming particles. According to the Seitz model \cite{ Seitz:1958nva}, the passage of incoming particles in the superheated liquid creates local agitation of the molecules, which transfer part of their energy to neighboring molecules. This phenomenon generates highly superheated heat-spike regions in which a small part of the liquid vaporizes, leading to the formation of a gaseous spherical cavity or protobubble within the active liquid itself. The expansion mechanism of this cavity is not specified further in the Seitz model. It only defines the limit beyond which the phenomenon becomes irreversible according to a critical radius, $r_c$, where the protobubble is in static equilibrium with the surrounding liquid. The critical radius $r_c$ in which the pressure differential across the surface is balanced by the surface tension is given by
\begin{equation}
    r_c=\frac{2\gamma}{P_b-P_l},
    \label{eq:critical_radius}
\end{equation}
where  $\gamma$ is the surface tension and $P_l$ and  $P_b$ are the pressure of the liquid and the bubble, respectively. If the radius of the protobubble is smaller than $r_c$, it will collapse, whereas if the radius is larger than $r_c$, it will undergo further expansion and form a macroscopic bubble.

Seitz \cite{Seitz:1958nva}, following Pless and Plano's work \cite{Pless:1956}, has defined the energy threshold or critical energy required to form a critically sized bubble as
    \begin{equation}
    \begin{aligned}
    Q_{\text{Seitz}} &= \frac{4 \pi}{3} r_c^3 \rho_b\Delta H+4 \pi r_c^2\left(\gamma -T_0 \frac{d \gamma}{d T}\right) \\
    &\quad -\frac{4}{3} \pi r_c^3\left(P_b-P_l\right)+W_{i r r},
    \end{aligned}
    \label{seitz}
    \end{equation}
    
\noindent where $\Delta H$ is the enthalpy of vaporization and $W_{irr}$ is the irreversible work term. $\rho_b$ is the vapor density of the protobubble which can be calculated by \cite{PICO:2019rsv}
    \begin{equation}
        \rho_b\approx\rho_v\frac{P_b}{P_v},
    \end{equation}
    and its pressure is given by
    \begin{equation}
        P_b  \approx P_v - \frac{\rho_v}{\rho_l}(P_v-P_l),
    \end{equation}
where  $P_v$ and $\rho_v$ are the saturated vapor pressure and density of the fluid at the detector operational temperature $T_0$, and $\rho_l$ represents the liquid density. The Seitz threshold $Q_{\text{Seitz}}$ is a thermodynamic property of the target fluid determined by two thermodynamic variables such as $T_0$ and $P_l$. This also allows us to replace one of the thermodynamic variables with $Q_{\text{Seitz}}$.

The critical energy takes into account three major components: the work done to expand a protobubble to the critical radius against the pressure of the surrounding liquid, the energy required to evaporate the liquid, and the work needed to form the protobubble surface (the liquid-vapor interface). The irreversible work $W_{irr}$ is usually neglected and includes the emission of acoustic waves, effects due to viscosity, energy lost to electronic excitation or ionization due to the recoil ion, and energy lost due to heat diffusion outside the critical radius.

To successfully form a macroscopic bubble in the Seitz model, an incoming particle has to deposit an energy, over a short distance, greater than the Seitz critical energy according to
    \begin{equation}
    E_{\mathrm{dep}} = \frac{dE_{\mathrm{dep}}}{dx} l_c = \frac{dE_{\mathrm{dep}}}{dx} b r_c > Q_{\text{Seitz}}.
    \label{eq:seitz_Edep}
    \end{equation}
    Here, $l_c$ represents the track length, also known as the critical length, and $b$ is a phenomenological parameter that varies in the literature \cite{Harper_1993,Denzel:2016epc}. For instance, in the PICO experiment \cite{Tardif:2019}, $b=2$ is considered for a C$_3$F$_8$ target. In our work, we consider the track length as a variable that depends on both the primary recoil energy and the probability of interaction with target nuclei.

\section{\label{sec:MD} Bubble formation: Molecular dynamics simulations}

The Seitz model describes the minimum energy required to form an initial bubble of radius $r_c$ and the conditions for the bubble to continuously grow and expand. However, the mechanism for forming the initial protobubble remains unclear. To address this, we use MD simulations to study the energy deposited through the nuclear recoil and the initial bubble growth. The MD simulation is performed using the Large-scale Atomic/Molecular Massively Parallel Simulator (\lammp{}) program \cite{Plimpton:2021}. In \lammp{}, we use the Lennard-Jones (LJ) potential to describe the interaction between atoms. In practice, to avoid unnecessarily heavy computation, we take advantage of the truncated shifted-force LJ potential, which can accurately describe the superheated system \cite{Denzel:2016epc, Kozynets:2019ihv} and provides a direct relationship between the simulation and the real temperature of the detector \cite{Errington:2003}.

In MD simulations, all quantities are unitless and are referred to as LJ units. They are expressed as multiples of some fundamental parameters $m$ (atomic mass), $\sigma$ (intermolecular distance), and $\epsilon$ (interaction energy), respectively, that depend on the target fluid under consideration. We also set the Boltzmann constant $k_B=1$. The transformation between \textit{Système International} and LJ units is shown in Table ~\ref{tab:lj_units}.

\begin{table}[ht]
\centering
\begin{tabular}{ccc}

Quantity & Variable  & LJ \\
\hline
Distance            & $x$     & $x^* = x/\sigma$   \\
Mass                & $M$  & $M^* = M/m$   \\
Time                & $\tau$     & $\tau^* = \tau\sqrt{\frac{\epsilon}{m\sigma^2}}$            \\
Energy              & $E$    & $E^* = E/\epsilon$           \\
Density             & $\rho$ & $\rho^* = \rho \frac{\sigma^3}{m}$ \\
Pressure            & $P$  & $P^* = P\frac{\sigma^3}{\epsilon}$ \\

\end{tabular}
\caption{Physical quantities expressed in LJ units.}
\label{tab:lj_units}
\end{table}

\lammp{} simulates the motion and interactions of atoms in a system using the LJ potential. In our work, we perform our simulation using periodic boundary conditions. To create a superheated state, we use the $NVE$ mode with constant particle number, volume, and energy. We use the superheated environment to simulate the energy deposition resulting from the track of an ionizing particle due to nuclear recoil. To model the track length, we defined it as a narrow cylinder with a length of $l_{c}$. The volume of the cylinder is calculated using the formula
\begin{equation}
V = \pi r_{\mathrm{cyl}}^2l_{c}.
\end{equation}
For all simulations, we fix the cylinder radius to be $r_{\mathrm{cyl}}=2\sigma$, as varying $r_{\mathrm{cyl}}$ between 2$\sigma$ and 5$\sigma$ does not affect bubble formation. To simulate the energy deposition event, we rescale the temperature of the target atoms inside the cylinder region by using the equipartition theorem as
\begin{equation}
T_{\mathrm{dep}} = \frac{2}{D}\frac{E_{\mathrm{dep}}}{N k_B}+T_0,
\label{eq:T_dep}
\end{equation}
where $D$ is the number of degrees of freedom of the target molecule or atom, $T_0$ is the temperature of the superheated liquid (which is also the temperature of the detector), and $N$ is the number of atoms or molecules inside the cylindrical region. The degree of freedom $D$ of a molecule depends on its structure. Noble elements with monoatomic structures, such as argon and xenon, have 3 translational degrees of freedom ($D=3$). In contrast, complex molecules like C$_3$F$_8$ and CF$_3$I have both 3 translational and 3 rotational degrees of freedom ($D=6$). 

Since the simulation box is a finite region of space, we need to avoid the influence of the bubble expansion on the overall pressure of the system. To solve this, we use the $NPT$ mode (constant number of particles, pressure, and temperature) to simulate the bubble growth. In this ensemble, the system can exchange energy and volume with external temperature and pressure baths. This approach is useful for investigating the behavior of liquids under constant pressure conditions. By maintaining a constant temperature and pressure, the $NPT$ ensemble provides a realistic representation of the thermodynamic properties and dynamics of the system to study phase transition phenomena. Through the simulation results, we can observe the growth or collapse of bubbles depending on the track length $l_{c}$ and energy deposition $E_{\rm dep}$. 

\begin{figure}[htbp]
\centering

\begin{subfigure}{.499\linewidth}
  \centering
  \includegraphics[width=\linewidth]{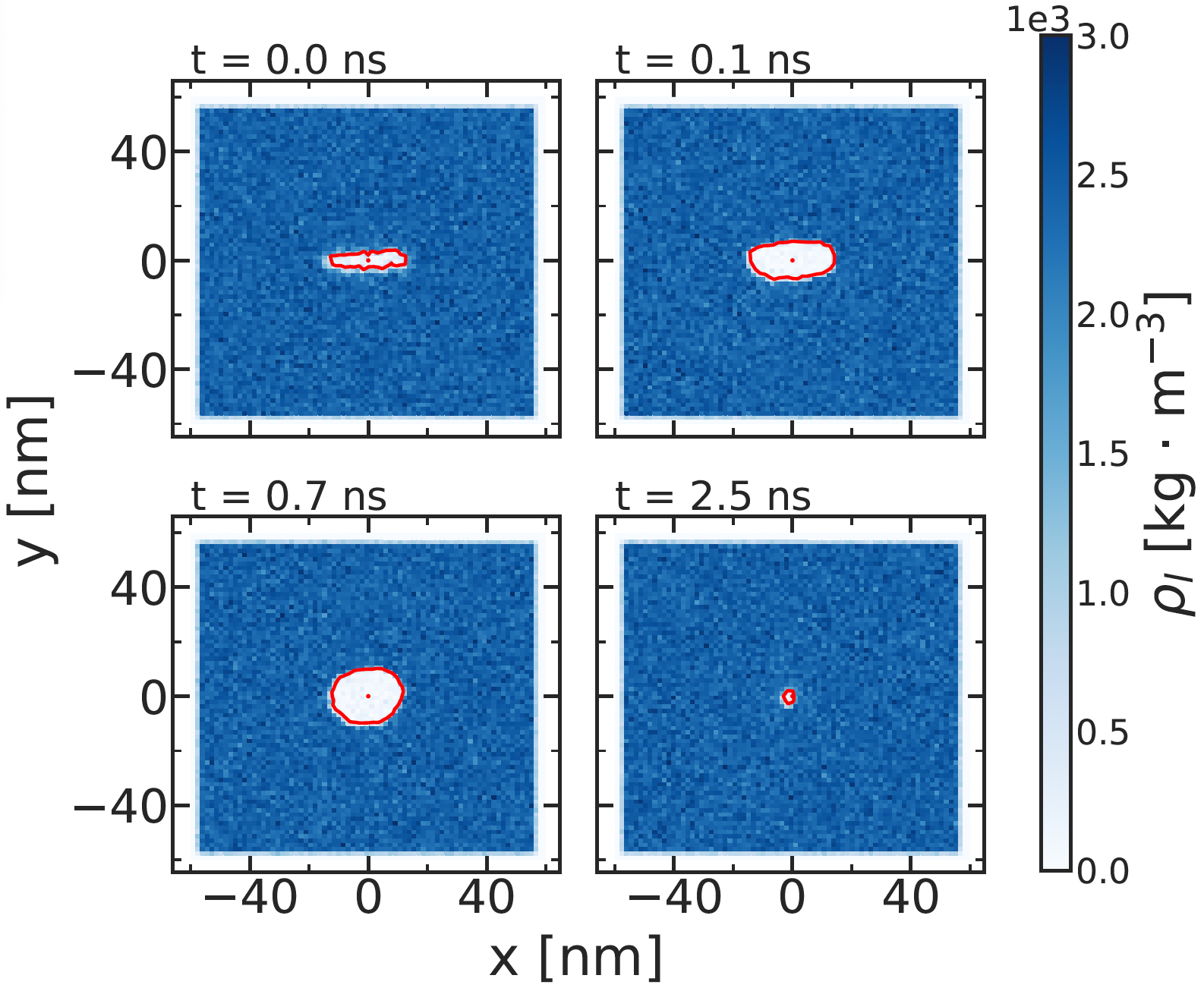}
  \label{fig:MD_xe1}
\end{subfigure}%
\hfill
\begin{subfigure}{.499\linewidth}
  \centering
  \includegraphics[width=\linewidth]{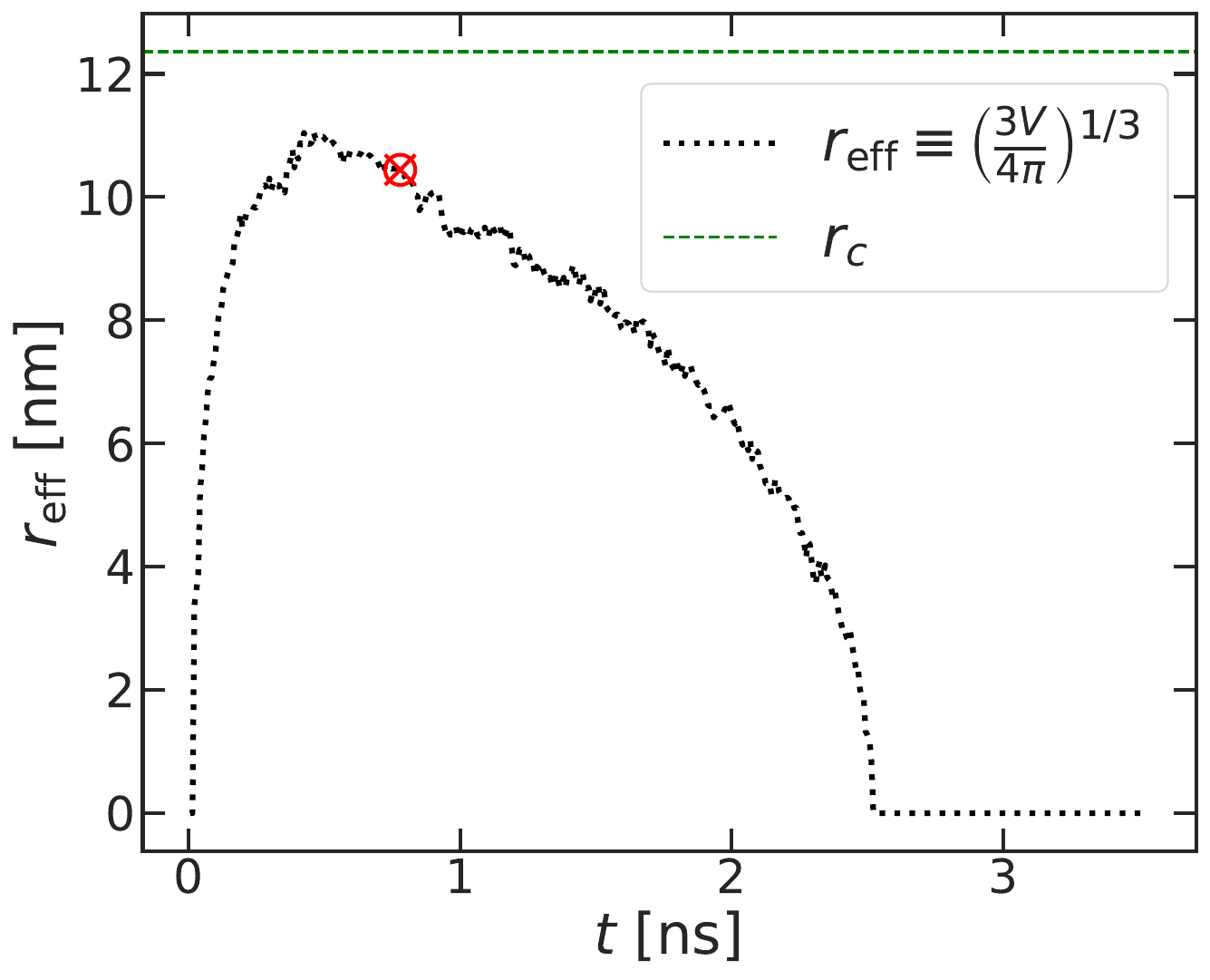}
  \label{fig:MD_xe2}
\end{subfigure}


\begin{subfigure}{.499\linewidth}
  \centering
  \includegraphics[width=\linewidth]{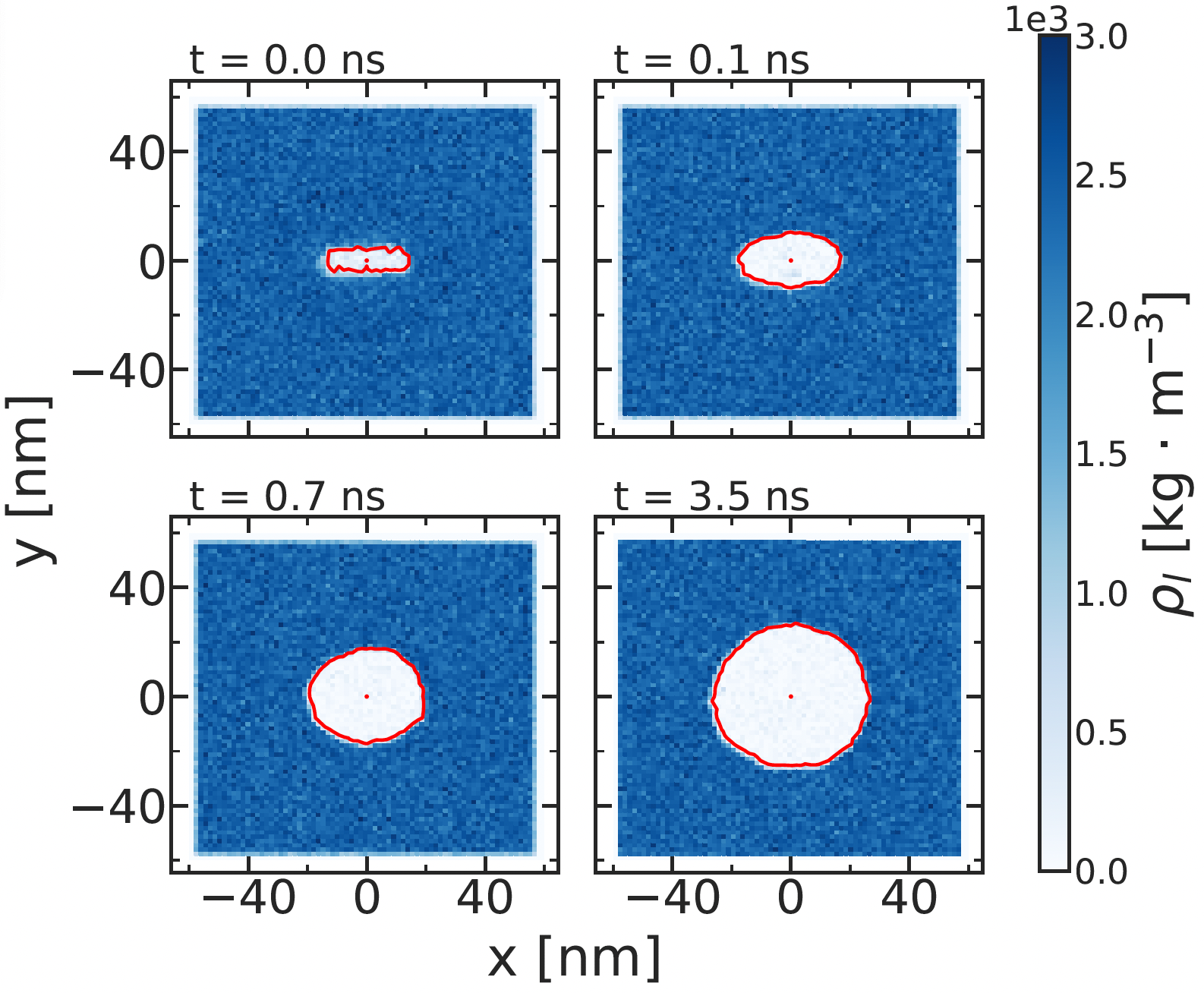}
  \label{fig:MD_xe3}
\end{subfigure}%
\hfill
\begin{subfigure}{.499\linewidth}
  \centering
  \includegraphics[width=\linewidth]{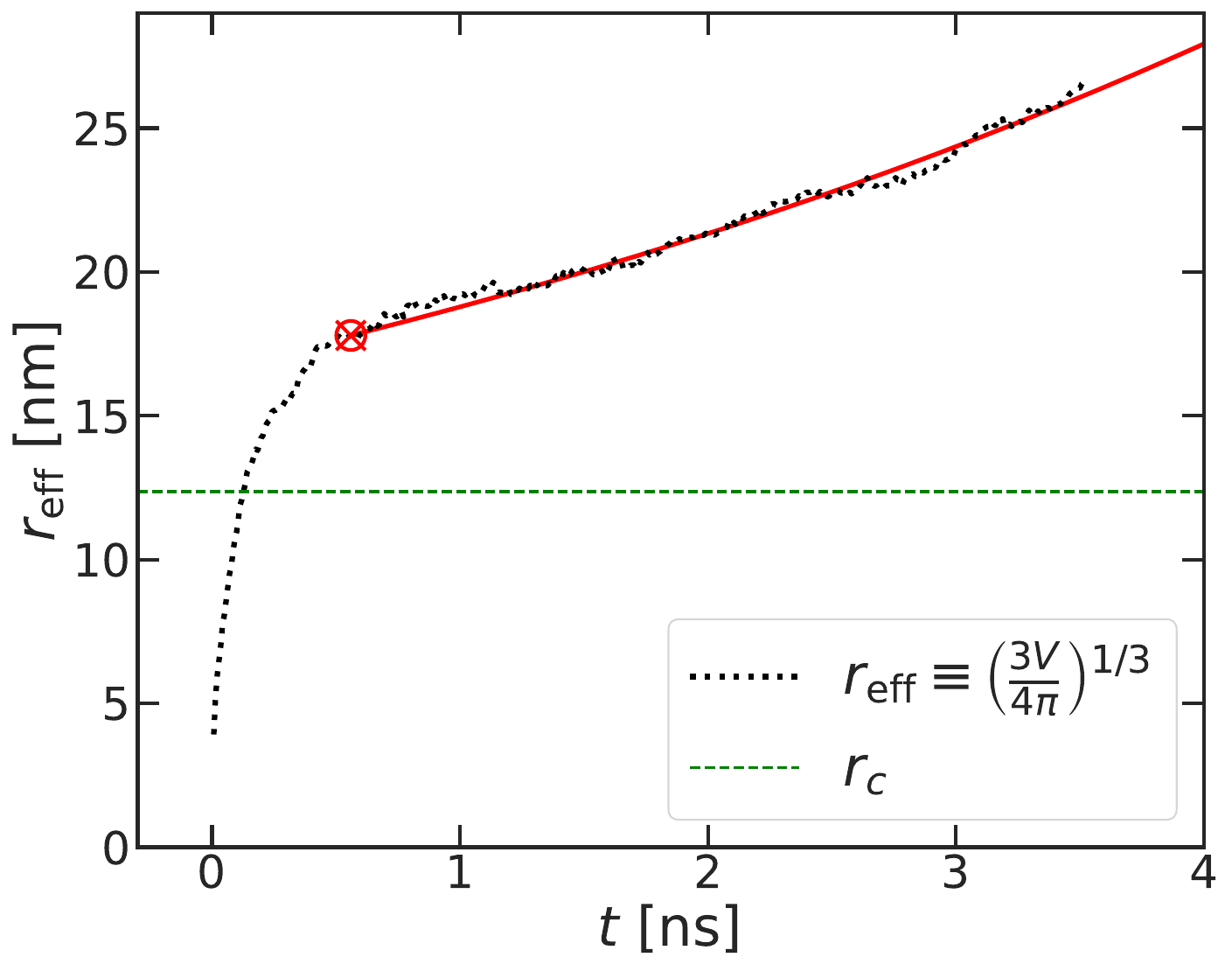}
  \label{fig:MD_xe4}
\end{subfigure}

\caption{MD simulation in liquid xenon at $Q_{\text{Seitz}} = 0.9$ keV for $E_{\text{dep}}=1$ keV (upper panels) and $E_{\text{dep}}=2$ keV (lower panels) with $l_c = 2r_c$. In the first case, no bubble is formed, while a bubble nucleates successfully in the second case, although both are above the Seitz threshold. The color scale shows the density of the liquid $\rho_{l}$ in the simulation region. The right panels show the effective radius of the bubble as a function of time. The red circle indicates the time when the bubble becomes spherical and the red line represents a fit of the bubble growth rate as in Refs. \cite{MIKIC1970657,Kozynets:2019ihv}. The green dashed line is the critical radius $r_c$ of liquid xenon obtained from Eq.~\eqref{eq:critical_radius}.}
\label{fig:MD_xe}
\end{figure}

Figure ~\ref{fig:MD_xe} displays two instances of energy deposition in our MD simulation for xenon with $E_{\rm dep}>Q_{\text{Seitz}}$. The \python{} package \verb|skimage.measure.find_contours| is used to track the surface of the spheres. On the right, the effective radius $r_{\rm eff} \equiv \sqrt[3]{3V/4\pi}$ represents the radius of a sphere that has the same volume as the nonspherical bubble. The red circle indicates the time when the bubble becomes spherical (eccentricity $<0.4$). Furthermore, the red line corresponds to a fit of the bubble growth rate to a combination of Rayleigh-Plesset \cite{rayleigh_1917} and Plesset-Zwick \cite{Pless:1956} equations as in \cite{MIKIC1970657,Kozynets:2019ihv}. It serves as a model for the growth of the bubble and it is in good agreement with our MD simulation. From the results of the simulation, we observe that even if the energy deposited is higher than the critical Seitz threshold, the bubble does not always form, which is compatible with what generally is observed in experiments. This finding is one of the main results of this work.

To investigate this further, we conduct MD simulations to study the formation and collapse of bubbles based on the variation of two key variables: the energy deposition $E_{\mathrm{dep}}$ and the cylinder length $l_c$, which also represents the track length. Our goal is to determine the conditions under which bubbles form or collapse, as illustrated in Fig.~\ref{fig:MD_xe}, to estimate the critical energy threshold. To analyze the results of bubble nucleation, we categorize them based on the linear energy density $dE_{\mathrm{dep}}/dx$ and the track length $l_c$ in the MD simulation. Following this procedure, we have conducted an analysis of various liquid species (C$_3$F$_8$, CF$_3$I, xenon and argon) with the Seitz threshold chosen as in bubble chamber experiments, enabling us to compare the results with existing data. Thanks to this method, we can deduce the actual energy threshold function of these target fluids from MD simulations. Theses results are summarized in Secs. \ref{sec:simuproc} and \ref{sec:simuprocB}.

\subsection{\label{sec:simuproc} Simulation setup}

To reproduce the thermodynamic parameters of the bubble chamber experiments in our MD simulations, we first determine the MD parameters in terms of LJ units. The data available from the PICO and SBC collaborations are summarized in Table~\ref{tab:gas_summary}. For argon, the SBC detector is planned to be operated at $Q_{\text{Seitz}} = 40$ eV but the projected cross-section limits are evaluated for 100 eV threshold \cite{Giampa:2021wte, Alfonso-Pita:2023frp}. For xenon, the temperature and pressure were measured only within a certain range. In MD simulations, the important parameters to reproduce the thermodynamic conditions are the temperature $T_0$ and the density of the liquid $\rho_l$ for each value of $Q_{\text{Seitz}}$. 


\begin{table}[h!]
\centering
\begin{tabular}{c*{8}{c}}
\hline\hline
Collaboration & Gas  & $Q_{\text{Seitz}}$  & Temperature $T_0$ & Pressure $P_l$\\
& & (keV) & ($^\circ$C) & (psia) \\
\hline
PICO   & C$_3$F$_8$  & 3.29 &  14 &  30 \\
       & C$_3$F$_8$  & 2.45 &  16 &  30  \\
       & CF$_3$I  & 13.6 &  34.5 &  23  \\
\hline
SBC   & Xe  & 0.9 &  --55 to --38 &  24 to 50  \\
       & Xe  & 1.48 & --55 to --38 &  24 to 50  \\
       & Xe & 2.0 &  --55 to --38 &  24 to 50  \\
Projected & Ar & 0.04 &  --143.15 &  20  \\

\hline\hline
\end{tabular}
\caption{Experimental thermodynamic conditions in several superheated liquids, available from the PICO \cite{PICO:2022nyi, Amole_2016} and SBC \cite{Durnford:2021cvb, alfonsopita2022snowmass} collaborations. For superheated argon, only the data from the projected sensitivity are available \cite{Giampa:2021wte}.}
\label{tab:gas_summary}
\end{table}

C$_3$F$_8$ and CF$_3$I are multiatom compounds, leading to a nonisotropic intermolecular potential. Considering the computational expense associated with accounting for all internal chemical bonds within the molecule, a degree of simplification is necessary for the simulation. This simplification is achieved through a method known as coarse graining (CG), which treats the molecule as a single particle while disregarding internal interactions, including chemical bonds. CG models are MD models specifically designed to investigate the behavior of complex molecules. They enable a reduction in system complexity by focusing on relevant physical quantities while still capturing the fundamental physics at play. 
For xenon and argon, we do not need to use the CG model because of the isotropic nature of the atom.

\begin{table}[h!]
\centering
\begin{tabular}{c*{8}{c}}
\hline\hline
 Gas & Temperature $T_c$   & Energy $\epsilon$  & Distance $\sigma$ & Mass $m$  \\
 &  (K) & (J) & ($\angstrom$) & (u) \\
 
\hline
C$_3$F$_8$  & 345.1 &  $5.095\times 10^{-21}$ &  5.38 & 188.02 \\
CF$_3$I  & 396.495 &  $5.85\times10^{-21}$ &  6.43 & 195.91 \\
\hline
Xe  & 289.74 &  $4.278\times 10^{-21}$ &  3.97 & 131.293\\
Ar & 150.86 & $2.227\times 10^{-21}$  &  3.4  & 39.948\\

\hline\hline
\end{tabular}
\caption{Parameters used to generate a superheated liquid state with MD simulations.}
\label{tab:MDValue_LJ}
\end{table}

To determine the parameters $\epsilon$ and $\sigma$ that correspond to a superheated liquid state, we first calculate the interaction energy parameter $\epsilon$ through the relation \cite{Errington:2003} $\epsilon = T_c k_B/0.935$, where $T_c$ represents the critical temperature of the gas, as determined by NIST \cite{NIST:FIPS1402}. For $\sigma$, we create a saturation state in the MD simulation to obtain the saturation liquid density $\rho^*_{\rm sat}$ and use \refprop{} \cite{LEMMON-RP10} to get the saturation  liquid density $\rho_{\rm sat}$, both values at $T_0$. In the case of xenon, we set the temperature to $ T_0 = -42.9^\circ$C. Then $\sigma$ follows from $\rho^*_{\rm sat}=\rho_{\rm sat}\sigma^3/m$. The values are shown in Table~\ref{tab:MDValue_LJ}, allowing us to express all the physical quantities in LJ units for our MD simulations. Table~\ref{tab:MDValue_summary} summarizes the values we  use in our MD simulations to reproduce the superheated state for each gas species and associated Seitz threshold $Q_{\rm Seitz}$. The pressure values $P_{\rm MD}$ represent the outputs obtained from the MD simulations.

\begin{table}[h!]
\centering
\begin{tabular}{c*{8}{c}}
\hline\hline
\multicolumn{2}{c}{$Q_{\text{Seitz}}$}& \multicolumn{2}{c}{ $T_0$}    & \multicolumn{2}{c}{$P_{\rm MD}$}  & \multicolumn{2}{c}{Density $\rho_l$} \\

Gas & (keV)   & ($^\circ$C) & ($\epsilon/k_B$)    &  (psia) &   ($\epsilon/\sigma^3$) & (g/cm$^3$) & ($m/\sigma^3$)  \\
\hline
C$_3$F$_8$& 3.29   & 14 & 0.778  & 106 &  0.0224  & 1.3676 &  0.688 \\
C$_3$F$_8$& 2.45   & 16 & 0.7834  & 71 &  0.0146 & 1.3791 &  0.672  \\
CF$_3$I& 13.6  & 34.5 & 0.725 & 31 &  0.0075  &  1.9797 &  0.7095\\
\hline
Xe& 0.9   & -42.9 & 0.743      & 25 &  0.00253  & 2.4204 &  0.6951 \\
Xe& 1.48   &  -42.9 & 0.743  & 56 &  0.00571 & 2.4237 &  0.6956  \\
Xe& 2.06    & -42.9 & 0.743  & 76 &  0.00767 & 2.4258 &  0.6959 \\
Ar& 0.0445   & -143.15 & 0.806     & 20 &  0.00247  & 1.035 &  0.639 \\

\hline\hline
\end{tabular}
\caption{Thermodynamic properties of the different gas species in both the \textit{Système International} and LJ units used in the MD simulation. The liquid density values $\rho_l$ are extracted from \refprop{} \cite{LEMMON-RP10} using the temperature $T_0$ and pressure $P_l$ from the experiments. The pressure values $P_{\rm MD}$ are the outputs obtained from the MD simulations.}
\label{tab:MDValue_summary}
\end{table}

As can be noticed in Table~\ref{tab:MDValue_summary}, for C$_3$F$_8$ and CF$_3$I, the use of a CG model, which involves approximating a compound molecule with a single isotropic particle, results in deviations from the experimental pressure conditions of the liquid. In the experiments, the temperature $T_0$ and pressure $P_{l}$ are controlled to achieve a superheated state. However, in the MD simulations, we regulate the temperature $T_0$ and the liquid density $\rho_l$ to attain the same state. When generating the superheated state in MD, the pressure becomes a dependent variable determined by temperature $T_0$ and liquid density $\rho_l$. Consequently, the CG model can lead to a discrepancy in the pressure estimation compared to the experiment. Despite this discrepancy, it is important to note that the CG model remains valid and has been employed successfully in past studies \cite{Kozynets:2019ihv, Denzel:2016epc}. Therefore, we constrained the model using real experimental temperature and liquid density values to conduct the simulations for C$_3$F$_8$ and CF$_3$I. 
For argon, the SBC Collaboration plans to operate the detector at $Q_{\text{Seitz}}=40$ eV \cite{Giampa:2021wte, Alfonso-Pita:2023frp}. Owing to the microfluctuations of pressure and liquid density in the MD simulations, the closest thermodynamic threshold value that we can achieve is $Q_{\text{Seitz}}=44.5$ eV, which is the one we use for our model.

\subsection{\label{sec:simuprocB} Energy threshold results from simulation}

\begin{figure}
    \centering
    \includegraphics[width = 1\columnwidth]{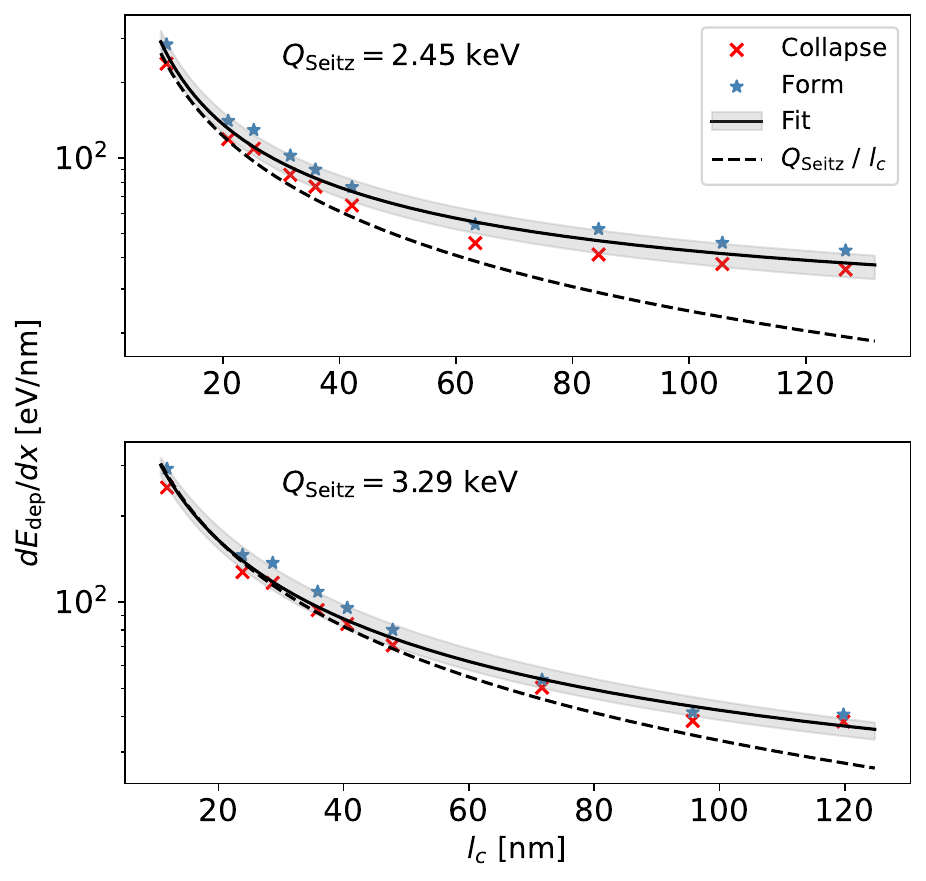}
    \caption{The bubble nucleation condition as a function of the linear energy density and the track length. Results are shown for C$_3$F$_8$ for $Q_{\text{Seitz}} =2.45$ keV and $Q_{\text{Seitz}} =3.29$ keV. The red crosses indicate the highest energy where bubble collapse was seen in simulation, while the blue stars indicate the lowest energy where bubbles were seen to form. The fitting curve is created using Eq.~\eqref{eq:fit_func} in between these limiting cases. The dashed line corresponds to the minimum energy threshold defined by the Seitz model $Q_{\text{Seitz}}$/$l_c$.}
    \label{fig:c3f8_bubble_curve}
\end{figure}

\begin{figure}
    \centering
    \includegraphics[width = 1\columnwidth]{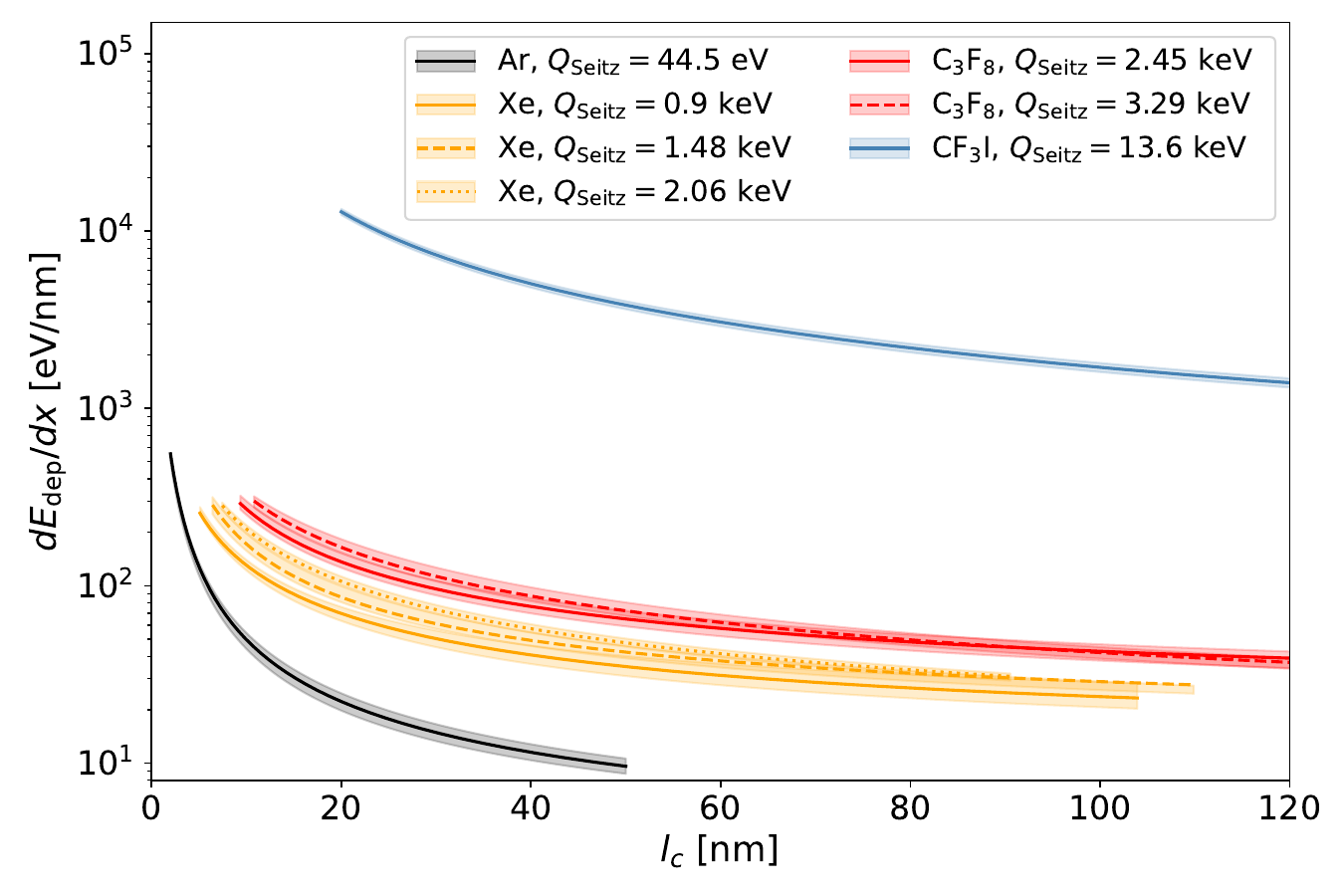}
    \caption{The bubble nucleation condition as a function of the linear energy density and the track length for each gas species and Seitz threshold available in experiments. }
    \label{fig:dedx_summary_plot}
\end{figure}

Once we created the superheated state corresponding to Seitz threshold values $Q_{\text{Seitz}}$, we vary the energy deposition $E_{\mathrm{dep}}$ and track length $l_c$ to determine whether the bubble forms or collapses. The results for C$_3$F$_8$ at $Q_{\text{Seitz}}=2.45$ keV and $Q_{\text{Seitz}}=3.29$ keV are summarized in Fig.~\ref{fig:c3f8_bubble_curve}. We use the function
\begin{equation}
    \frac{dE_{\mathrm{dep}}}{dx}   = \frac{E_{\mathrm{dep}}}{l_c} = \frac{a}{{l_c}^2}+\frac{b}{l_c}+c
    \label{eq:fit_func}
\end{equation}
to fit the region between bubble formation and collapse. The parameters $a$, $b$, and $c$ are free parameters of the fit, which are determined based on the simulation data. The lower and upper bounds for $a$, $b$, and $c$ are determined by fitting the largest simulated energy density resulting in bubble collapse and the lowest energy density resulting in bubble formation, respectively. We compare our results with the Seitz model by computing $Q_{\text{Seitz}}$/$l_c$, which is found to be below our prediction, as expected. As a note: $l_c=2r_c$, as used in the PICO experiment for C$_3$F$_8$, would correspond to about $50$ nm. We follow the same procedure to determine the energy threshold function for each gas species. The results are summarized in Fig.~\ref{fig:dedx_summary_plot}. These fit functions represent the energy threshold function of our model and are used to calculate the nucleation efficiency. The fit parameters with errors for all gas species are given in Appendix \ref{sec:fit_param}.

\section{\label{sec:MCLind} Recoil ion in target: Monte Carlo simulation}

MD simulations provide insights into the mechanism of bubble nucleation. The initial trigger for nucleation is the deposition of energy, typically in the form of heat, known as the heat spike. The rise in temperature of the target fluid is due to momentum transfer from the incident particle during nuclear recoil events, which is ionized and dissipates energy in the process. 

SRIM and its Monte Carlo (MC) subprogram, transport of ions in matter (TRIM), are simulation tools used to model interactions between ions and their target material \cite{Ziegler:2010}. They can be used to study the behavior of ions when they collide with solid or liquid targets. The important parameter to set the threshold conditions is the density of the liquid $\rho_l$ for each target material that can be found in Table \ref{tab:MDValue_summary}. In SRIM/TRIM MC simulations, the incident ion is assigned a specific energy $E_{\text{dep}}$ and direction, and the simulation calculates the ion trajectory as it traverses the target material. Throughout its journey, the ion undergoes collisions with atoms in the target, resulting in a cascade of recoiling atoms. The SRIM/TRIM simulation uses universal nuclear-scattering cross sections, as provided by \cite{Ziegler:2010}, which quantitatively measure the probability of a collision leading to a recoil event. The output includes the energy and position of the recoiling atoms and the track length of the ion. 

In our model, we use this information to calculate the stopping power, which consists of two mechanisms. On the one hand, energy dissipation of the ion can occur via electronic excitation or further ionization caused by interactions with the electron clouds surrounding the target atoms, and we usually call this the electronic stopping power. Ultimately, this energy deposition contributes to scintillation or fluorescence, depending on the medium. On the other hand, another portion of the energy is dissipated through momentum transfer via elastic collisions with other nuclei in the material, and we usually call this the nuclear stopping power. These collisions prompt additional ionization, and the resulting primary ions create cascades of successive interactions. This momentum transfer from recoiling contributes to the generation of a localized heat spike. This energy loss is commonly described by Lindhard's model \cite{Lindhard:1961zz} and the fraction of nuclear recoil energy contributing to the electronic excitation is called the Lindhard or quenching factor $\mathcal{L}$. The numerical determination of the Lindhard factor is given in Appendix \ref{sec:Lindhard_F}.

 In our MC simulation, we send ions with an assigned energy $E_{\text{dep}}$ into the target. Each incoming ion undergoes multiple elastic scattering events with the target nuclei, which in turn leads to the creation of secondary recoiling ions. Between each recoiling event, the initial ion gradually loses energy as it interacts with the electron clouds surrounding the target atoms, i.e. electronic stopping. The secondary ions, generated by the incident nucleus, also experience energy loss along their trajectories following the same principle.  However, this portion of the nuclear stopping energy is subsequently carried by the secondary ions, which can also dissipate energy through electronic stopping as they travel, and this information is not provided by the SRIM/TRIM simulation. To include it in our model, we use the Lindhard model to implement corrections that account for the energy losses incurred during the process of electronic stopping. When considering the energy of the secondary ion with index $i$ in a nuclear recoil event, we can describe it through
\begin{equation}
    E_{i, n} = E_i (1-\mathcal{L}(E_i)),
    \label{eq:Ein}
\end{equation}
where $E_{i, n}$ represents the nuclear stopping energy loss experienced by the secondary ion with index $i$, $E_i$ is the initial energy of the secondary ion and $\mathcal{L}$ is the theoretical Lindhard factor. The recoil energy caused by the incident ion, $E_i$, is the calculation output of the MC simulation. Reference \cite{Hitachi:2008kf} approximates the Lindhard factor of a compound with the linear combination of $\mathcal{L}$ for each element, such as for $\mathrm{C_3F_8}$, $\mathcal{L}(\mathrm{C_3F_8}) = \left(3\mathcal{L}(\mathrm{C})+8\mathcal{L}(\mathrm{F})\right)/11 $. For the case of noble liquids, such as Xe and Ar, we use the noble element simulation technique (\nest{}) \cite{NEST:2023} package to determine the Lindhard factor. \nest{} fits the empirical functions from various experimental data. 

The Lindhard factor is a function that encapsulates the dependency of energy loss on nuclear recoil energy. Since the Lindhard factor $\mathcal{L}$ represents the fraction of energy loss due to electronic stopping, the remaining loss contributes to nuclear stopping. If we sum up all of the secondary ions that lost energy through nuclear stopping, we get the amount of energy that would contribute to the heat spike as 
\begin{equation}
Q = \sum_i\left(1-\mathcal{L}(E_i)\right)E_i.
\label{eq:lindhard_heat}
\end{equation}
Furthermore, the linear energy density is given by
\begin{equation}
\frac{dE_{\mathrm{dep}}}{dx}=\frac{Q}{l_c},
\label{eq:avg_dedx}
\end{equation}

\noindent where $Q$ is the energy loss in heat in Eq.~\eqref{eq:lindhard_heat} and $l_c$ is the track length varying for different events, determined from the ion trajectory in the MC simulation. 

\section{\label{sec:NuEff} Nucleation efficiency}

The key element of our approach is to include the Lindhard factor correction in the MC simulation for each secondary ion produced and to compare the results with the energy threshold functions obtained from MD simulations (referred to as the MD curve). In this way, we can determine if a given event from the MC simulation satisfies the condition necessary for bubble formation. If the energy density in an event is above the MD curve, the formation of a bubble is considered to occur. Conversely, if the energy density is below the MD curve, the bubble is considered to collapse.

\begin{figure}
  \centering
  \includegraphics[width=\columnwidth]{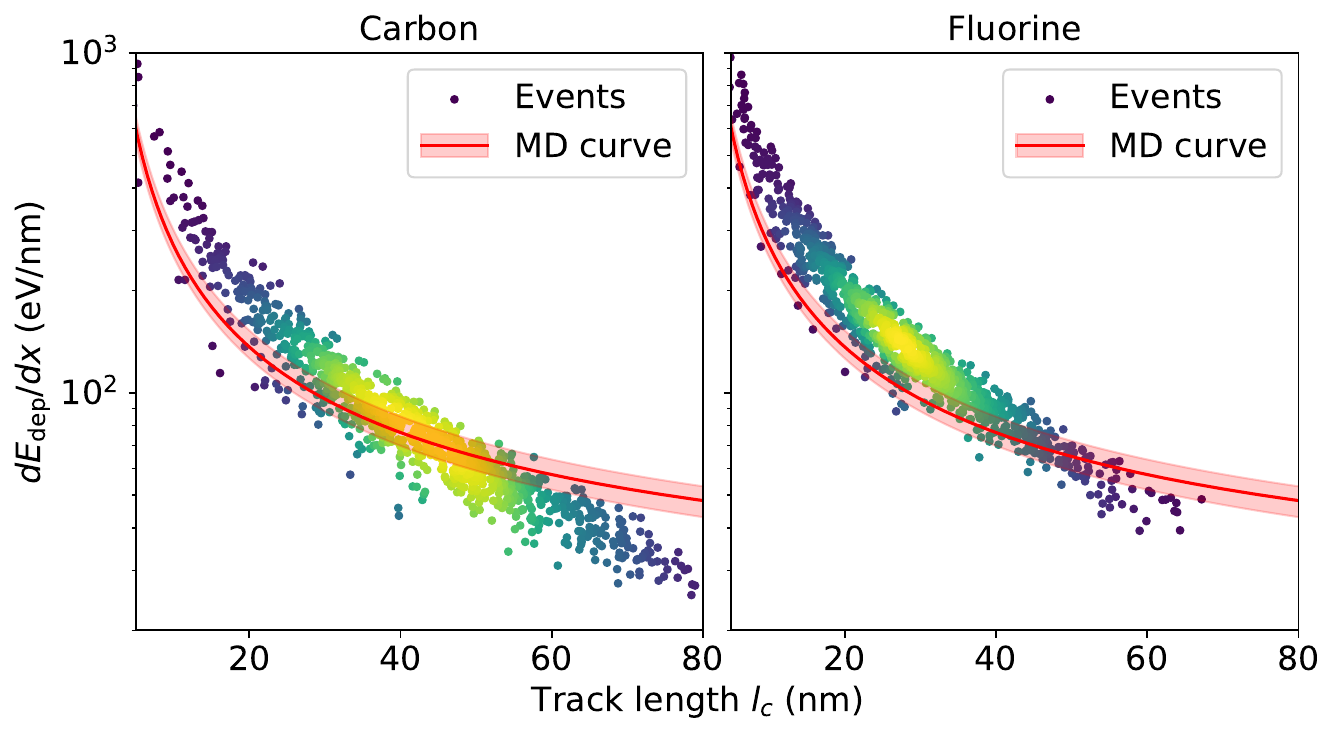}
    \caption{Example with $E_{\text{dep}}=6$ keV nuclear recoil events for C$_3$F$_8$ at $Q_{\text{Seitz}}=3.29$ keV. The $y$ axis represents the linear energy density and the $x$ axis represents the track length. The data points are the simulated events produced in MC simulation, including the Lindhard factor correction. The red curve represents the linear energy density fit using Eq.~\ref{eq:fit_func} from MD simulation at $Q_{\text{Seitz}}=3.29$ along with its uncertainty.  Left panel: the carbon recoiling events. Right panel: the fluorine recoiling events.  }
    \label{fig:CF_differ}
\end{figure}

An example of the combined results is presented in  Fig. \ref{fig:CF_differ} for an $E_{\text{dep}}=6$ keV nuclear recoil event in C$_3$F$_8$ for carbon and fluorine at $Q_{\text{Seitz}}=3.29$ keV. Although both carbon and fluorine ions possess the same initial energy, fluorine ions have a higher stopping power, enabling them to be stopped within a shorter distance. Thus, fluorine ions locally result in a higher energy deposition density than carbon ions. Moreover, it is observed that fluorine ions generate a greater number of events above the MD curve, indicating that bubble formation is more readily achieved with fluorine recoil. 

The nucleation efficiency $\varepsilon$ is calculated using
\begin{equation}
    \varepsilon = \frac{\text{events above the MD curve}}{\text{ total events}}.
    \label{eq:nuc_eff}
\end{equation}

\noindent For $E_{\rm dep}=6$ keV carbon and fluorine recoiling ions in C$_3$F$_8$, 47\% of carbon recoil events and 94\% of fluorine recoil events exceed the MD curve. Therefore, the bubble nucleation efficiency for carbon and fluorine recoil is 47\% and 94\%, respectively. Using this principle, we can determine the nucleation efficiency as a function of nuclear recoil energy and compare this model with experimental results.

\subsection{Nucleation efficiency calculation and uncertainties}
\begin{figure}
    \centering
    \includegraphics[width=\columnwidth]{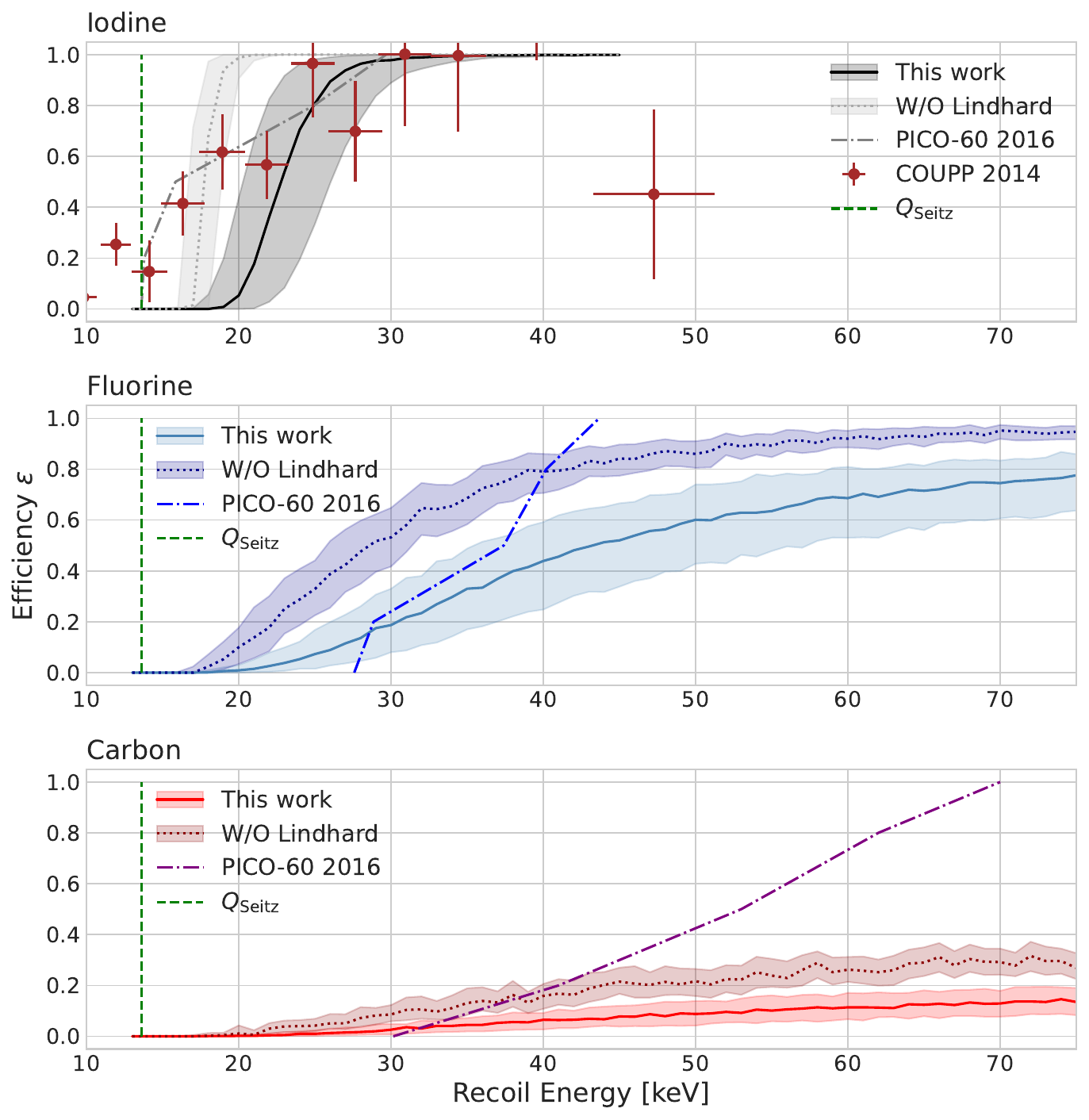}
    \caption{ Nucleation efficiency of CF$_3$I at $Q_{\text{Seitz}}=13.6$ keV indicated by the green vertical dashed line. The black, blue, and red solid curves represent the nucleation efficiency of iodide, fluorine, and carbon nuclei, respectively, obtained through calculations in this study with the Lindhard factor correction. The dotted curve represents our model without including the Lindhard factor correction. The dash-dotted gray, dark blue, and purple curves represent the experimental results for iodide, fluorine, and carbon recoil, respectively, given in \cite{Amole_2016}. The brown data points correspond to the iodine nucleation efficiency as a function of iodine-equivalent recoil energy from the COUPP experiment \cite{COUPP:2013yjn}.}
    \label{fig:eff_cf3i_srim}
    \end{figure}

We calculate the nucleation efficiency for the different targets through MD simulation in \lammp{}, and MC simulation in SRIM/TRIM with and without the secondary recoiling correction using the Lindhard factor. We then compare our results with data available for C$_3$F$_8$ in Fig.~\ref{fig:eff_c3f8_srim}, CF$_3$I in Fig.~\ref{fig:eff_cf3i_srim} and xenon in Fig.~\ref{fig:xe_eff_srim}. For argon, since there is no data available yet, we give a prediction for the nucleation efficiency at $Q_{\text{Seitz}}=44.5$ eV set by the experiment \cite{Giampa:2021wte, Alfonso-Pita:2023frp}. 
For all cases, the error estimation in our calculation incorporates both the upper and lower limits of the function defined in Eq.~\eqref{eq:fit_func} governing the formation of bubbles. The upper and lower limits for $\varepsilon$ are obtained from the upper and lower limits on $a, b,$ and $c$ in the fit. We also take into account the error on the $k$ parameter of the Lindhard factor (see Appendix \ref{sec:Lindhard_F}). Additionally, there is a component of statistical error resulting from the inherent randomness of the MC simulations. To quantify this statistical uncertainty, we employ the bootstrap method. We find that the magnitude of this statistical uncertainty is small compared to the error arising from the upper and lower limits.

\subsection{Nucleation efficiency for C$_3$F$_8$}

\begin{figure}
    \centering
    \includegraphics[width=\columnwidth]{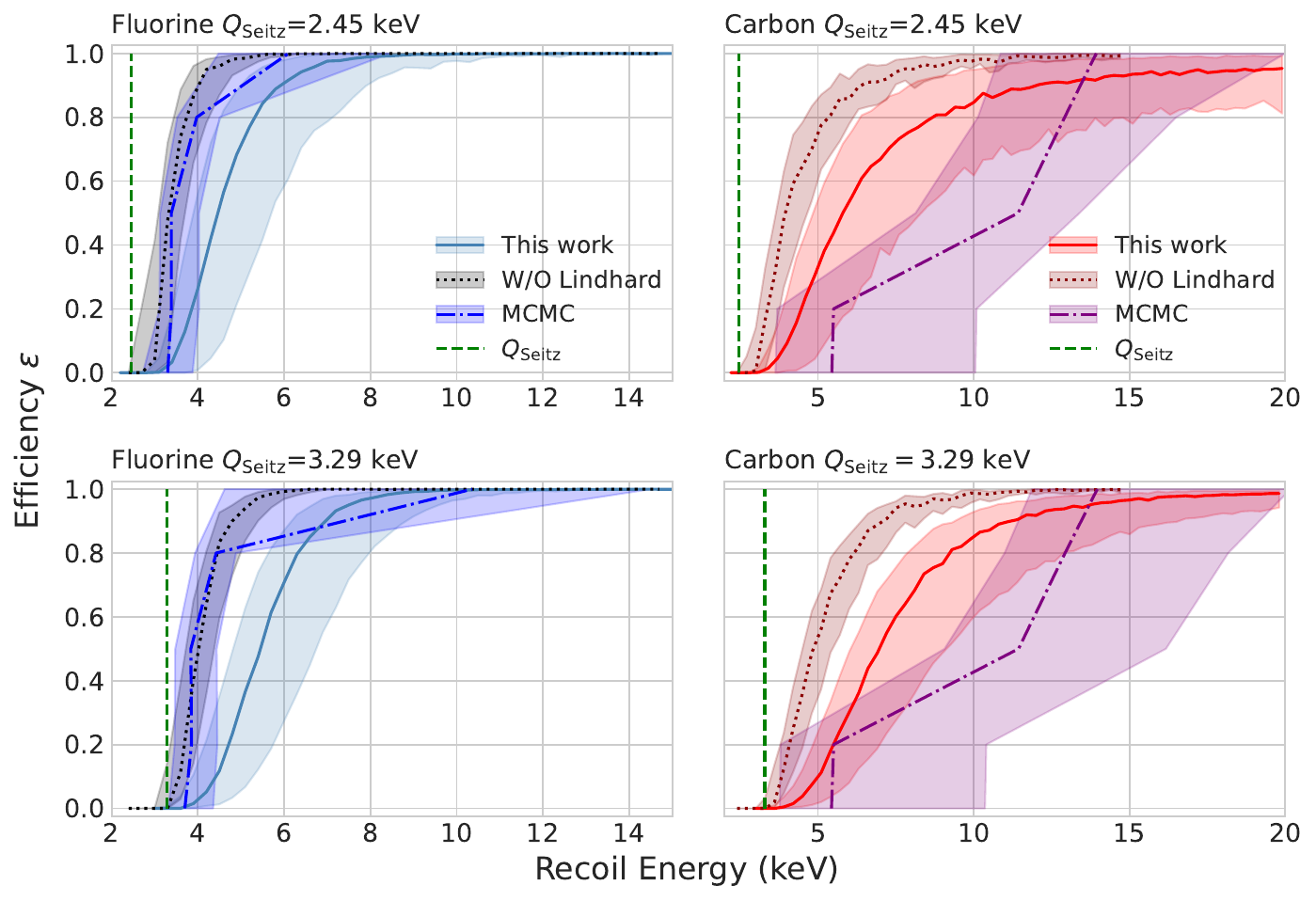}
    \caption{Nucleation efficiencies of C$_3$F$_8$ at Seitz thresholds of 3.29 and 2.45 keV (dashed green vertical line), respectively. The recoil energy corresponds to $E_{\rm dep}$. The blue curve shows the nucleation efficiency of fluorine nuclei, while the red curve shows the one for carbon; both are obtained with our model (MD, MC, and Lindhard factor) along with their uncertainties. The dash-dotted dark blue and purple curves are the experimental results for fluorine and carbon recoils, respectively, obtained with MCMC in \cite{PICO:2022nyi}. The dotted curves along with the uncertainties represent our model without including the Lindhard factor correction. }
    \label{fig:eff_c3f8_srim}
    \end{figure}

Figure \ref{fig:eff_c3f8_srim} presents the nucleation efficiency for C$_3$F$_8$ from our model with and without the Lindhard factor compared to the best-fit experimental data obtained via Markov chain Monte Carlo (MCMC) analysis of the PICO-60 experiment \cite{PICO:2022nyi}. The recoil energy corresponds to $E_{\rm dep}$.
With the Lindhard factor, for $Q_\text{Seitz}=3.29$ keV, the fluorine efficiency from our model aligns well with the experimental analysis within the low- and high-efficiency regions, falling within a 1$\sigma$ error band. However, it is observed to be approximately 1.5$\sigma$ higher than the experimental analysis near $60\%$ efficiency. For $Q_\text{Seitz}=2.45$ keV, the fluorine efficiency curves obtained from both the experiment and our work reveal good agreement within a 1$\sigma$ error band. Regarding the carbon simulation, it matches the experimental analysis in both low- and high-efficiency regions. However, there is a deviation between $30\%$ and $90\%$ efficiency, with an energy approximately 1$\sigma$ to 2$\sigma$ lower than the experimental value. It is interesting to note that the fluorine efficiency without the Lindhard factor correction falls within the 1$\sigma$ error band of the experimental data, indicating a very good agreement. In contrast, the efficiency for carbon with the Lindhard factor correction aligns more closely with the experimental analysis.

\subsection{Nucleation efficiency for CF$_3$I}

Following the same procedure, we calculate the nucleation efficiency including the errors for CF$_3$I, as depicted in Fig.~\ref{fig:eff_cf3i_srim}. We compare our results to experimental data obtained through neutron calibration from the COUPP  \cite{COUPP:2013yjn} and PICO-60 experiments \cite{Amole_2016}. We observe that the iodine recoil simulations exhibit a close alignment with the experimental results obtained from both the PICO-60 experiment (gray curve) and the COUPP iodine efficiency data (brown data points) within a 1$\sigma$ confidence interval for efficiencies exceeding 60\%. However, below 60\% efficiency, there is a notable discrepancy between the iodine simulation results and the experimental data. Regarding the fluorine recoil, the experimental results demonstrate agreement with the simulation when the efficiency is below 50\%. However, as the energy increases, the efficiency of the simulation begins to plateau, deviating from the experimental trend. The case of carbon recoil shows a slower increase in efficiency compared to the experimental result, indicating a disagreement between the two.


\begin{figure}
    \centering
    \includegraphics[width=\columnwidth]
    {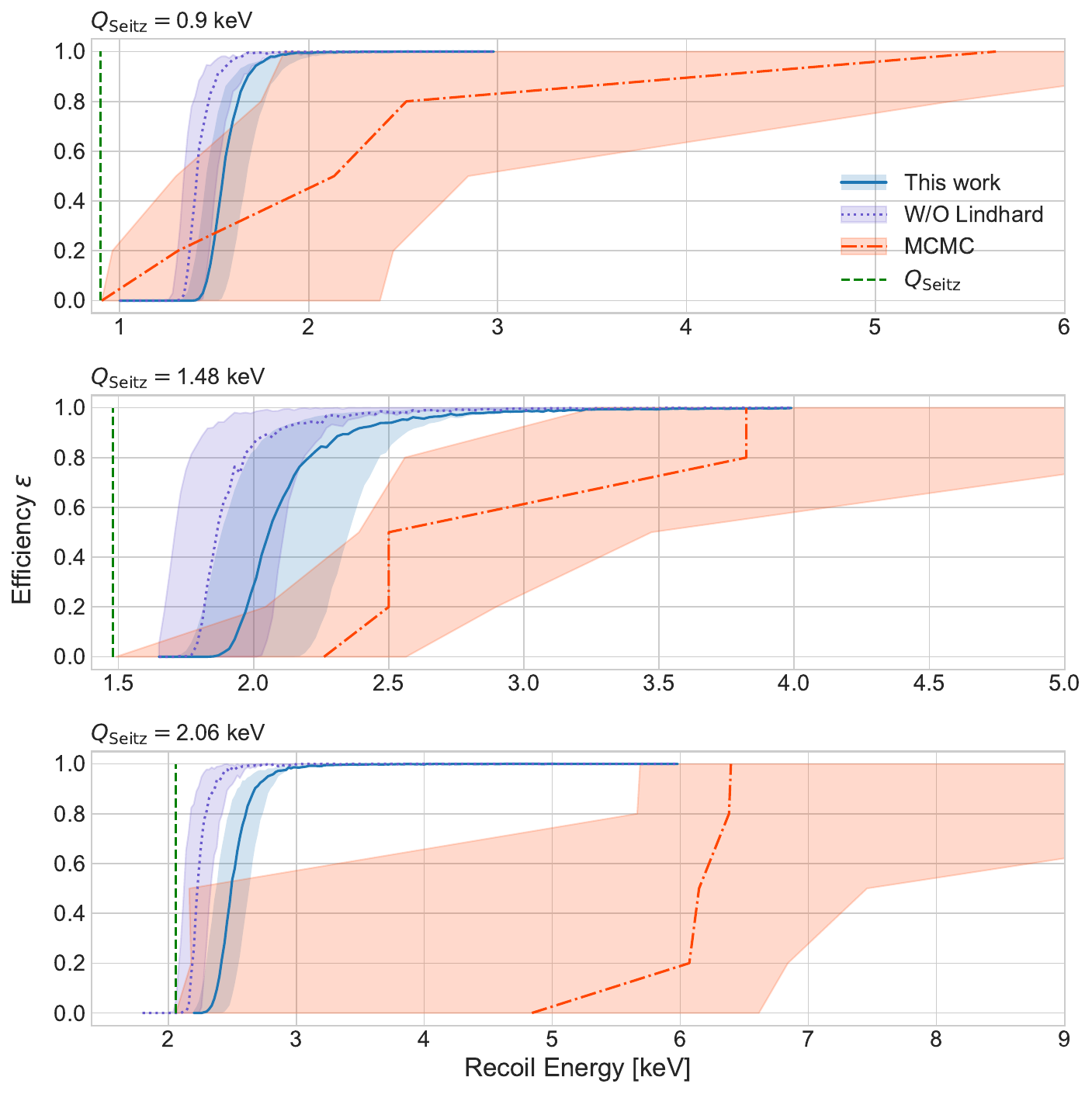}
    \caption{Xenon bubble nucleation efficiency as a function of energy $Q_{\text{Seitz}}$ (the green dashed lines) of 0.9, 1.48, and 2.06 keV, respectively. The blue curve is the result from the present work with MD and MC simulation and the Lindhard correction along with uncertainties. The dotted curve and uncertainties represent our model without the  Lindhard factor correction. The orange-red dash-dotted curve is obtained from experimental data via MCMC at 0.9, 1.48 \cite{alfonsopita2022snowmass}, and 2 keV threshold \cite{Durnford:2021cvb}.}
    \label{fig:xe_eff_srim}
\end{figure}

\subsection{Nucleation efficiency for xenon}

Figure~\ref{fig:xe_eff_srim} shows the xenon nucleation efficiency with Seitz thresholds of 0.9, 1.48, and 2.06 keV, respectively.  The experimental best fit is done with the MCMC approach found in \cite{Durnford:2021cvb, alfonsopita2022snowmass}. For the 0.9 keV threshold condition, the curve calculated using our model mostly agrees with the experimental result within the 1$\sigma$ error band. However, it should be noted that for the 1.48 and 2.06 keV threshold condition, the uncertainty of the results in this study appears to be underestimated, particularly in the high-efficiency regions. 

While the errors in MCMC are not explicitly mentioned in any publications cited, it is worth noting that the MCMC results are poorly constrained at high efficiency due to limited statistics of high-energy recoils during the neutron calibration. The approximate method used to plot the 1$\sigma$ uncertainty bands in the MCMC may not reflect the resulting asymmetric uncertainties accurately. The likelihood function was assumed to be Gaussian, which is not the case. This approximate treatment of uncertainties that are actually asymmetric and non-Gaussian may misrepresent the agreement, or lack thereof, between the MCMC results and our current work. The same applies to the MCMC results for C$_3$F$_8$ but in this case, the uncertainties are in general smaller because of the higher statistics in the neutron calibration data. Nonetheless, within the 1$\sigma$ error band, the experimental curve and the simulation show agreement for efficiencies at least up to 50\%. Above this threshold, the experimental data for 2.06 keV reveals a sharp increase in the energy required to nucleate the bubble, exceeding 10 keV. Higher statistics would be needed to gain a deeper understanding of the underlying physics governing these phenomena.

\subsection{Results discussion}

In general, the results obtained from our model reproduce the nucleation efficiency obtained by the experiments in large parts qualitatively and even quantitatively. At least 50\% of our simulation results are consistent within 1$\sigma$ when compared to existing experimental data. The main systematic uncertainties come from the lack of information in the Lindhard model at low energy [O(eV)] and the CG approximation used in the MD simulations for fluorocarbons. In our study, we employ the Lindhard model to estimate the energy loss of secondary ions generated by the initial recoiling ion. As a result, the Lindhard factor correction shifts the nucleation efficiency curves to higher recoil energy, which aligns with experimental observations. For the case of xenon, this correction seems to have a stronger impact to reproduce the nucleation efficiency data compared, while for C$_3$F$_8$ and CF$_3$I, it appears to overestimate the energy threshold for fluorine and iodine, respectively.

A possible explanation for a better agreement with noble liquids could be that the Lindhard factor obtained from \nest{} benefits from the numerous experimental data available. In contrast, for C$_3$F$_8$ and CF$_3$I, the Lindhard factor is solely calculated based on theoretical grounds, lacking experimental verification (see Appendix \ref{sec:Lindhard_F}). It should also be noted that for any Seitz threshold energy in our work [O(keV)], the secondary ions'energy is lower than 1 keV, and more on the order of O(eV). Even though we have \nest{} and a great amount of data for noble gases such as xenon and argon, there are limited data below 1 keV. These results indicate the potential of using bubble nucleation efficiency to constrain the Lindhard factor for low-energy nuclear recoils where tensions exist between current experiments. Furthermore, the MD simulation employs a simplified CG model, treating C$_3$F$_8$ and CF$_3$I as a sphere, while the MC simulation does not account for molecular structure. Assuming the validity of the Lindhard correction, the discrepancy may stem from the influence of the structure of the molecule. Similar considerations apply to CF$_3$I, where the complexity of the molecule and the uncertainty associated with the Lindhard correction introduce challenges. The endeavor to construct a full molecular simulation incorporating chemical bonds has been contemplated. However, it is important to note that a fully detailed MD simulation without CG is computationally demanding and would not yield results within a reasonable time frame. The use of neural networks and machine learning techniques is a promising way to enhance the accuracy of the MD simulation \cite{ML_MD,AI_MD}. Overall, the efficiency results for the primary targets, fluorine in C$_3$F$_8$ and iodine in CF$_3$I, are reasonably accurate within 1$\sigma$ even without incorporating experimental data.

\begin{figure}
        \centering
        \includegraphics[width=0.9\columnwidth]{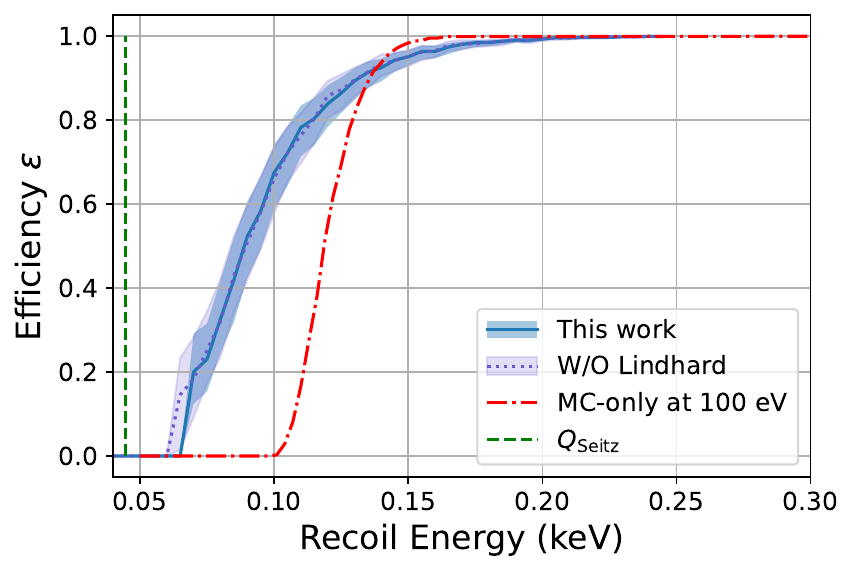}
        \caption{Predicted nucleation efficiency of liquid argon at $Q_{\text{Seitz}}=44.5$ eV (the dashed green line) for the future SBC experiment. The blue and dotted curves are derived from our model (MD and MC simulations) with and without the Lindhard factor correction, respectively, along with uncertainties. The dash-dotted red curve is obtained with MC-only simulation for $Q_{\text{Seitz}}=100$ eV. }
        \label{fig:eff_ar}
    \end{figure}

\subsection{Nucleation efficiency prediction for argon}

In the upcoming SBC experiment \cite{Giampa:2021wte, Alfonso-Pita:2023frp}, the plan is to utilize superheated liquid argon for the dark matter search. To prepare for this future operation, we predict from our model the nucleation efficiency for $Q_{\text{Seitz}}=44.5$ eV. The results are presented in Fig. \ref{fig:eff_ar} with and without the Lindhard factor correction (obtained from \nest{}). It is interesting to note that since the Lindhard factor is minimal for argon at low energy, it has almost no impact on the nucleation efficiency. In addition, we also compare our results with the curve obtained from MC-only simulations at $Q_{\text{Seitz}}=100$ eV. This is the threshold used for the SBC projected sensitivity \cite{Giampa:2021wte, Alfonso-Pita:2023frp}. Once data will be available, our predictions can be tested against experimental results.
\\
\\


\section{\label{sec:CS_DM} Impact on exclusion curves for dark matter}

The nucleation efficiency plays a crucial role in determining the exclusion limits for dark matter in bubble chamber detectors, making it a significant factor to consider. Understanding the underlying principles that govern nucleation efficiency can provide valuable insights for planning future experiments. The event rate for a given WIMP mass $m_\chi$ is obtained from
\begin{equation}
R_{\rm obs} (m_\chi) =\int_{E_{\min }}^{E_{\max }} \varepsilon\left(E_{\rm nr}(m_\chi)\right) \frac{d R}{d E_{\rm nr}} d E_{\rm nr},
\label{eq:R_obs6}
\end{equation}
where $R_{\rm obs}$ is the observed event rate, $\varepsilon$ is the nucleation efficiency, $E_{\rm min}$ and $E_{\rm max}$ are the lower and upper energy limits of region of interest, and $dR/dE_{\rm nr}$ is the differential rate. Using Eq. \eqref{eq:R_obs6}, we can determine the observed event rate of WIMPs for a given mass ($m_\chi$) and directly include our results for the nucleation efficiency. The 90\% confidence level (CL) upper limits of the cross section in the spin-dependent (SD) and spin-independent (SI) sectors are calculated with the profiled Feldman and Cousins \cite{Feldman_1998} method. The exclusion limit $\sigma_{\lim}$ for a given mass of WIMP $m_\chi$ is expressed as
\begin{equation}
    \sigma_{\lim}(m_\chi) = n_{\lim} \frac{\sigma(m_\chi)}{R_{\rm obs}(m_\chi)\cdot \tau},
    \label{eq:excl_lim}
\end{equation}
where $\tau$ is the exposure with the dimension of time multiplied by mass. $n_{\lim}$ is the 90$\%$ upper limit event in Table IV of Ref. \cite{Feldman_1998} and depends on the experimental event counts and prior background counts. The cross section $\sigma$ and differential rate $dR/dE_{\rm nr}$ are extracted from \verb|dmdd| package \cite{dmdd_Gluscevic_2015}. We use standard WIMP halo parameters also chosen by the experiments \cite{Lewin:1995rx} such as local dark matter density $\rho_{\rm D}= 0.3$ GeV/c$^{-2}$cm$^{-3}$, galactic escape velocity $v_{\rm esc} = 544$ km/s, the velocity of Earth with respect to the halo $v_{\rm Earth} = 232$ km/s, and the characteristic WIMP velocity with respect to the halo $v_0 = 220$ km/s.

\begin{figure}
    \begin{subfigure}{0.99\linewidth}
        \includegraphics[width=\linewidth]{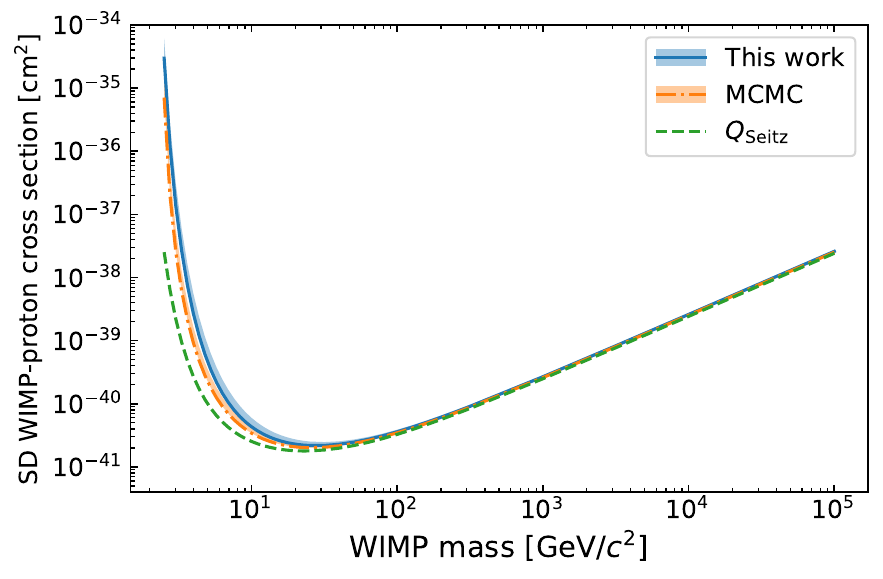}
        
        \label{fig:limit_c3f8_sd}
    \end{subfigure}
    \begin{subfigure}{0.99\linewidth}
    \includegraphics[width=\columnwidth]{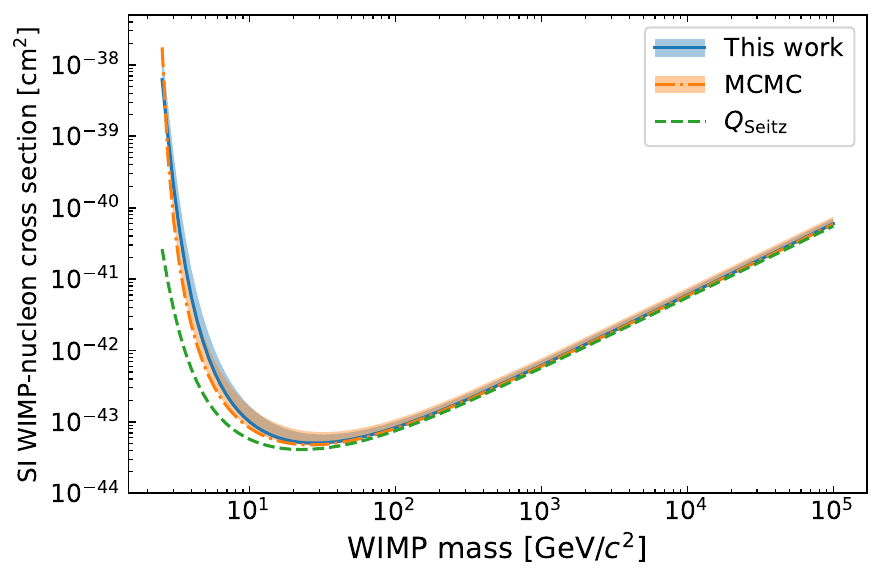}
    \label{fig:limit_c3f8_si}
    \end{subfigure} 

    \caption{The 90\% CL cross-section limits for C$_3$F$_8$ under conditions used in the PICO-60 2019 complete exposure \cite{PICO:2019vsc} in the SD (top panel) and SI (bottom panel) sectors. The blue curve is calculated with our efficiency function, while the orange dash-dotted curve is derived from the MCMC efficiency function in \cite{Durnford:2021cvb}. The shaded region represents the 1$\sigma$ efficiency upper limit area propagated to the cross section. The green dashed line is the efficiency obtained from the Seitz model. }
    \label{fig:limit_c3f8}
\end{figure}

\begin{figure}
    \begin{subfigure}{0.99\linewidth}
    \includegraphics[width=\linewidth]{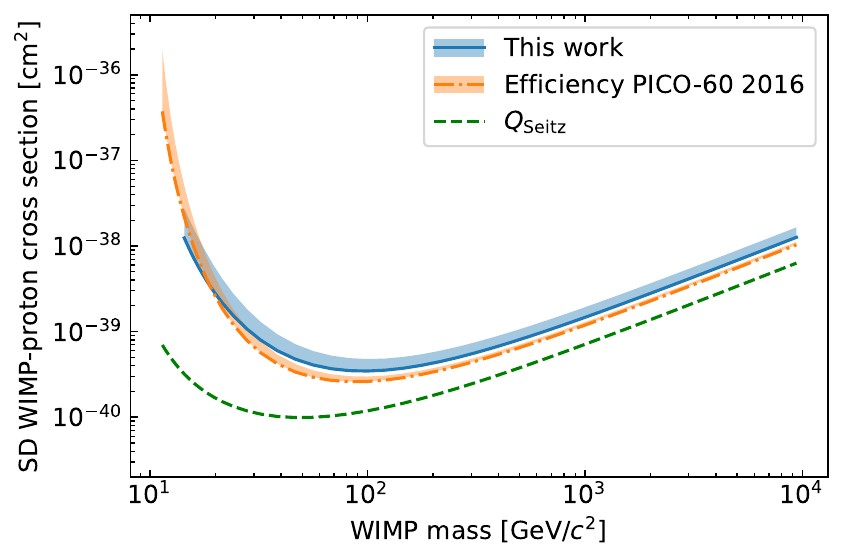}

        \label{fig:limit_cf3i_sd}
           \end{subfigure}
    \begin{subfigure}{0.99\linewidth}
        \includegraphics[width=\columnwidth]{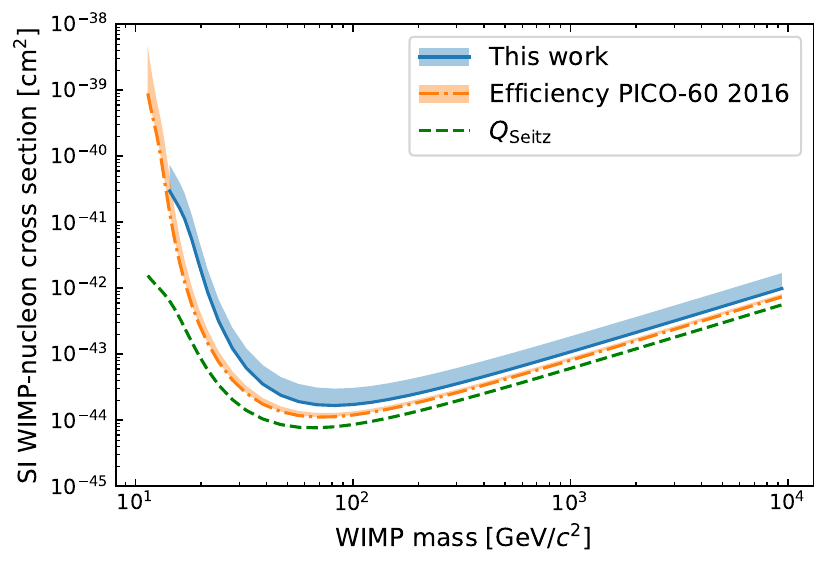}

    \label{fig:limit_cf3i_si}
    \end{subfigure} 
  
    \caption{Comparison of SD (top panel) and SI (bottom panel) exclusion limits for CF$_3$I at $Q_{\rm Seitz}=13.6$ keV as relevant for PICO-60 2016 \cite{Amole_2016}. The blue curve is calculated with our model-generated efficiency, while the orange dash-dotted curve is derived from the efficiency best-fit function in \cite{Amole_2016} and associated uncertainties. The green dashed line is the efficiency obtained from the Seitz model.  }
    \label{fig:limit_cf3i}
\end{figure}

By using this method, we calculate the SD and SI cross-section limits for dark matter. We use the same exposure and observed event rates as in \cite{PICO:2019vsc} for C$_3$F$_8$ and in \cite{Amole_2016} for CF$_3$I. The observed event rate from our model is obtained with our nucleation efficiency including the Lindhard factor, which is integrated with the differential rate in Eq.~\eqref{eq:R_obs6}. The uncertainty is calculated by propagating the 1$\sigma$ uncertainty of the efficiency. We employ the same method to calculate the cross-section limits from other nucleation efficiency functions. To compare this with the Seitz model prediction, we use a step function starting at the $Q_{\rm Seitz}$ used for each target. 

 Figure~\ref{fig:limit_c3f8} shows our results compared with the efficiency obtained from MCMC \cite{Durnford:2021cvb} and the prediction with the Seitz model for C$_3$F$_8$ target at $Q_{\rm Seitz}=2.45$ keV and $Q_{\rm Seitz}=3.29$ keV. The exclusion curves differ only in utilizing different efficiency functions. Efficiencies, controlled by the energy threshold, affect the exclusion curves for low WIMP masses at the limit of sensitivity. As can be seen, in the high WIMP mass range, the cross-section limit curves are mostly aligned. This is because a higher WIMP mass corresponds to a higher recoil energy, and the nucleation efficiency of C$_3$F$_8$ is 100\% for high energy. However, discrepancies begin to emerge around the curve minima and in the lower mass range due to the variation in efficiency (see Fig. \ref{fig:eff_c3f8_srim}). The discrepancy between our model and the MCMC efficiency functions is within an order of magnitude. In the context of SI cross sections, the small discrepancy can be attributed to the fact that the simulated carbon and fluorine efficiencies both lie between the efficiencies obtained from the MCMC analysis. This intermediate positioning mitigates some of the discrepancies between the two. On the other hand, in the SD cross section, where only the fluorine nucleus contributes, the exclusion limit produced by our work appears to be more conservative. This is due to the simulated efficiency being slightly lower than the efficiency derived from the MCMC analysis. The discrepancy is even higher when compared with the Seitz prediction, as expected, which underestimates the energy threshold and where the efficiency of nuclear recoils is assumed to be a step function.

For CF$_3$I, Fig.~\ref{fig:limit_cf3i} shows the 90\% CL upper limit cross section from our model, from the efficiency function obtained by PICO-60 2016 \cite{Amole_2016} and the Seitz model prediction for $Q_{\rm Seitz}=13.6$ keV. In this case, in the high WIMP mass region, we observe that the exclusion curves slightly deviate from each other since the efficiency curves do not reach 100\% for fluorine and especially carbon (Fig. \ref{fig:eff_cf3i_srim}). In the SI cross section, there is a relatively high discrepancy between the results in this work and in the PICO-60 2016 efficiency function for the low WIMP mass range. This is due mainly to the carbon nucleation efficiency, which contributes mostly to the low-mass range and is lower in our simulated results. On the other hand, for the SD cross section, the results are mostly consistent within the same order of magnitude due to only fluorine and iodine contributing to the SD cross section. As observed previously, the discrepancy is even higher when compared with the Seitz prediction.

\begin{figure}
        \centering
        \includegraphics[width=0.99\columnwidth]{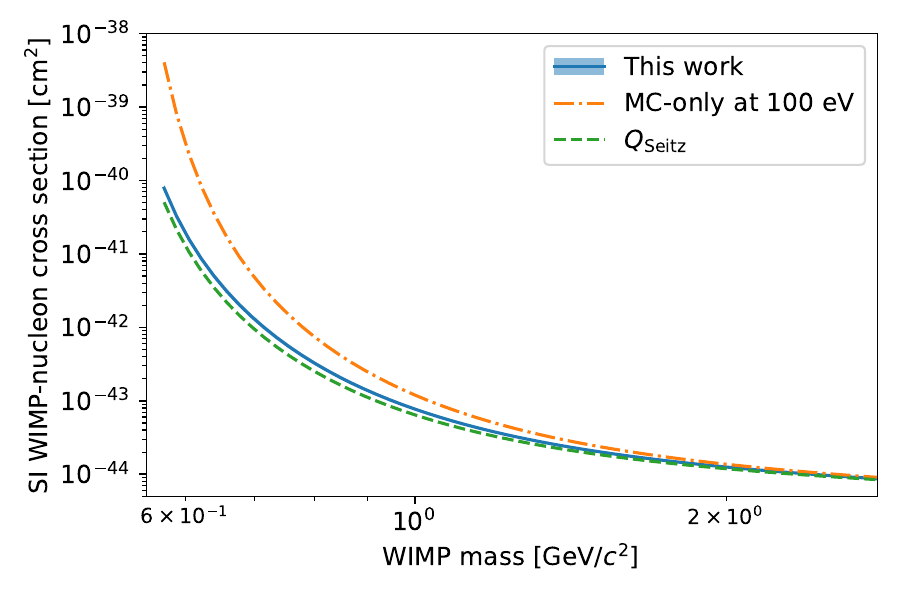}
        \caption{Predicted 90\% CL for the future SBC experiment, with the efficiency derived for $Q_{\rm Seitz}=44.5$ eV from our work (the blue line) and the Seitz model (the green dashed line). The uncertainty from our work is too small to be seen in the plot. The orange dash-dotted line is obtained from the efficiency from MC-only simulation for 100 eV threshold to compare with the projected sensitivity of the SBC detector \cite{Giampa:2021wte, Alfonso-Pita:2023frp}.}
        \label{fig:eff_ar_limit}
    \end{figure}

For argon, we estimate the 90\% CL cross-section limits in the SI sector shown in Fig.\ref{fig:eff_ar_limit} using the same exposure and observed event rate predicted in \cite{Giampa:2021wte, Alfonso-Pita:2023frp}. We compare our result to the Seitz model prediction at $Q_{\rm Seitz}=44.5$ eV. The exclusion curve predicted with our model shifts the sensitivity to higher WIMP masses as expected from the nucleation efficiency function obtained (Fig. \ref{fig:eff_ar}). We also present the limits obtained by the efficiency from MC-only simulation with SRIM/TRIM for $Q_{\rm Seitz}=100$ eV, which does not include MD and Lindhard corrections. The latter is used as a comparison since a 100 eV threshold is used for the projected sensitivity of the SBC detector. This is more conservative than our prediction. It will be interesting to compare it with experimental results when they are available in the future.

Using our model with MD and MC simulations including the Lindhard factor shows promise in assessing both bubble nucleation efficiency and estimating cross-section limits with an accuracy within the same order of magnitude as obtained by the PICO experiment with the C$_3$F$_8$ target. With these results, including our cross-section prediction for an argon target, we show that our model can be used as an accurate prediction of future dark matter projection limits for the PICO and SBC collaborations. The comparison with the exclusion curves obtained with the Seitz model clearly demonstrates that it overestimates the sensitivity of the experiments. In summary, this methodological approach provides valuable insight into understanding the intrinsic constraints of detectors. It further advances our understanding of their operational limitations and assists with experimental operation planning.

\section{Summary and conclusion}

Our approach begins by using molecular dynamics simulations to investigate the formation of bubbles at the microscopic scale.  
The simulations involve depositing energy in a cylinder to simulate the track length. Depending on the energy and length of the cylinder, the bubbles will either grow or collapse. 
An empirical function (linear energy density as a function of track length) is used to fit the events between bubble formation and collapse. The resulting empirical function is used as an acceptance or rejection criterion to determine the bubble nucleation efficiency in the Monte Carlo simulation as a function of recoil energy.

In addition, Monte Carlo simulations are employed to investigate the transfer of energy from recoiling nuclei to the target material. The energy loss during this process can be divided into nuclear stopping and electronic stopping. However, only the energy attributed to nuclear stopping contributes to the heat necessary for bubble formation. The Lindhard factor is used to describe the energy transfer and approximate the energy loss of secondary ions. 

By combining molecular dynamics and Monte Carlo simulations including the Lindhard factor correction, it is possible to determine the efficiency of bubble nucleation. The criterion for successful bubble formation is satisfied when the linear energy density exceeds the threshold defined by the empirical function. Conversely, if the condition is not met, the bubble will collapse. Simulations were conducted across various nuclear recoil energies using  C$_3$F$_8$, CF$_3$I, xenon, and argon to determine the relationship between bubble nucleation efficiency and nuclear recoil. The results show promising agreement in at least 50\% of our simulation with experimental data, thus surpassing the current Seitz model. The main systematic uncertainties come from the lack of information in the Lindhard model at low energy [O(eV)] and the CG approximation used in the MD simulations for fluorocarbons. To further improve our model over the entire range of the recoil energy, (1) dedicated measurements of the Lindard factor would be required at low energy, (2) more neutron calibration data with xenon and argon is needed to increase the statistics, and (3) one should simulate the molecule as a whole instead of using the CG approximation. For item (3), the use of neural networks or machine learning techniques \cite{ML_MD,AI_MD} is a promising direction. It is worth noting that our model can also be applied to estimating the nucleation efficiency for future target fluids where no experimental data exist.

Using our nucleation efficiency model enables us to reproduce cross-section limits. The results allow for comparison with experimental observations, facilitating a comprehensive assessment of how the simulated outcomes impact the exclusion limit. Efficiency affects the low-mass region, and the exclusion limits reproduced in this work agree with the published results within 1 order of magnitude or less for the PICO experiment with C$_3$F$_8$. For CF$_3$I, the agreement is consistent within 1$\sigma$ and 2$\sigma$ for spin-dependent and spin-independent channels, respectively.

Our methodology not only offers insights and predictions for the planning of future experiments but also has the potential to promote further understanding of the microphysics in bubble chambers at low-energy recoils. In addition, the use of our model can constrain the efficiency uncertainty from experimental calibration results in bubble chambers. Our work also reveals the potential of bubble nucleation efficiency to determine the Lindhard factor or nuclear quenching factor at low energy.

\begin{acknowledgments}
This research was undertaken thanks to the Canada First Research Excellence Fund through the Arthur B. McDonald Canadian Astroparticle Physics Research Institute. We gratefully acknowledge the support provided by Compute Ontario \cite{ontario}, the BC DRI Group, and the Digital Research Alliance of Canada \cite{alliance} required to undertake part of the simulation of this work. We thank Tetiana Kozynets for help with the MD simulations in the initial stage of the project.  We also thank Alejandro Santiago Garcia Viltres for sharing the cross-section limit calculation for the PICO-60 2019 results. In addition, we wish to acknowledge Daniel Durnford for a fruitful discussion regarding the MCMC uncertainties.

\end{acknowledgments}

\appendix

\section{ \label{sec:fit_param}\textsc{Fit parameters for energy threshold functions}}
In our model, the condition of nucleation is defined by the fitting function in Eq.~\eqref{eq:fit_func}, where $a$, $b$, and $c$ are free parameters of the fit based on the MD simulation data for each gas species. The results are summarized in Table \ref{tab:all_abc} including the upper and lower  bounds for the uncertainities.

\begin{table}[ht]
\centering
\begin{tabular}{cccc}
\hline\hline
Seitz threshold & a (eV$\cdot$nm) & b (eV) & c (eV/nm) \\ \hline\hline
 &  C$_3$F$_8$ &  \\
\hline
2.45 keV &  1543 &  2975.6 &  12.054 \\
2.45 keV lower bound & 112.48 &  2866.1 &  10.137 \\
2.45 keV upper bound &  32.204 &    3333 &  13.003 \\ \hline
3.29 keV & 3850.8 &  2111.9 &  21.169 \\
3.29 keV lower bound & 3131 &  2022.8 &  17.278 \\
3.29 keV upper bound & 3532.1 &  2433.8 &  22.084 \\ \hline
\hline
 &  CF$_3$I &  \\
\hline
13.6 keV &  2211 & 1454 & 28.066 \\
13.6 keV lower bound & 2211 & 1354 & 28.066 \\
13.6 keV upper bound & 2211 & 1554 & 28.066 \\\hline
\hline
 &  Xenon &  \\
\hline
0.9 keV & 705.86 & 1109.1 & 12.542 \\
0.9 keV lower bound & 1188.4 & 995.18 & 10.581 \\
0.9 keV upper bound & 1087.6 & 1116.3 & 17.417 \\ \hline
1.48 keV & 3113.4 & 1247.8 & 16.013 \\
1.48 keV lower bound & 1793.3 & 1264.9 & 12.969 \\
1.48 keV upper bound & 1991.8 & 1646.0 & 12.217 \\ \hline
2.06 keV & 1503.4 & 1835.2 & 10.371 \\
2.06 keV lower bound & 1388.4 & 1755.1 & 9.6365 \\
2.06 keV upper bound & 721.58 & 2031.3 & 9.4678 \\ \hline\hline
 &  Argon &  \\
\hline
0.0445 keV &  15.834 & 31.109 & 2.7416 \\
0.0445 keV lower bound & 14.617 & 26.638 & 2.7928 \\
0.0445 keV upper bound & 14.770 & 36.878 & 2.6474 \\ \hline\hline

\end{tabular}
\caption{Bubble nucleation fitting parameters in Eq.~\eqref{eq:fit_func} for each gas species and threshold.  }
\label{tab:all_abc}
\end{table}

\section{ \label{sec:Lindhard_F}\textsc{Lindhard Factor}}
The fraction of nuclear recoil energy contributing to the electronic excitation is called the Lindhard or quenching factor. We approximate the Lindhard factor by
\begin{equation}
\mathcal{L}= \frac{E_e}{E_n + E_e} \approx\frac{k g(\hat{E})}{1+k g(\hat{E})},
\label{lindhard_analytic}
\end{equation}
where
$$
g(\hat{E})=3 \hat{E}^{0.15}+0.7 \hat{E}^{0.6}+\hat{E}.
$$
Here, $E_n$ is the energy loss in nuclear stopping, $E_e$ is the energy loss in electronic stopping, $\hat{E}$ is the dimensionless reduced energy, and $k$ is a parameter associated with the electronic stopping power. Reference \cite{Hitachi:2008kf} shows that for most cases $k = 0.1-0.2$, and calculates  $k_\mathrm{C} = 0.127$ for carbon and  $k_\mathrm{F} = 0.132$ for fluorine in CF$_4$ by using
\begin{equation}
k = 0.133 Z^{2/3}A^{-1/2},
\end{equation}
where $A$ is the mass number. In our study, we are using a similar $k$ value for C$_3$F$_8$ that we vary between 0.1 and 0.2 to estimate the uncertainty, since it is made of the same atoms. For CF$_3$I, we choose $k_\mathrm{I} = 0.15$ and use 0.1 and 0.2 as bounds for estimating the uncertainty. The reduced energy can be expressed as
\begin{equation}
\hat{E}=\frac{a_{\mathrm{s}} A_2}{Z_1 Z_2 e^2\left(A_1+A_2\right)} E,
\end{equation}
where $a_s$ is the screening radius proposed by Lindhard \cite{Lindhard:1963},
\begin{equation}
a_{\mathrm{s}}=0.8853 a_0 /\left(Z_1^{2 / 3}+Z_2^{2 / 3}\right)^{1 / 2},
\end{equation}
where $Z$ is the atomic number, $a_0$ is the Bohr radius, subscript 1 represents the incoming ion, and subscript 2 represents the target medium atoms. If the target and projectile are the same elements ($Z_1=Z_2 = Z$), we have $\hat{E}=11.5\left(E/\text{keV}\right) Z^{(-7 / 3)}$. In this work, for argon and xenon, we use the Lindhard factor calculated in \nest{} \cite{NEST:2023}, as it provides the most up-to-date experimental fitting for the Lindhard factor \cite{Szydagis:2021hfh}.

\bibliography{apssamp_v9_corr}

\begin{thebibliography}{41}%
\makeatletter
\providecommand \@ifxundefined [1]{%
 \@ifx{#1\undefined}
}%
\providecommand \@ifnum [1]{%
 \ifnum #1\expandafter \@firstoftwo
 \else \expandafter \@secondoftwo
 \fi
}%
\providecommand \@ifx [1]{%
 \ifx #1\expandafter \@firstoftwo
 \else \expandafter \@secondoftwo
 \fi
}%
\providecommand \natexlab [1]{#1}%
\providecommand \enquote  [1]{``#1''}%
\providecommand \bibnamefont  [1]{#1}%
\providecommand \bibfnamefont [1]{#1}%
\providecommand \citenamefont [1]{#1}%
\providecommand \href@noop [0]{\@secondoftwo}%
\providecommand \href [0]{\begingroup \@sanitize@url \@href}%
\providecommand \@href[1]{\@@startlink{#1}\@@href}%
\providecommand \@@href[1]{\endgroup#1\@@endlink}%
\providecommand \@sanitize@url [0]{\catcode `\\12\catcode `\$12\catcode `\&12\catcode `\#12\catcode `\^12\catcode `\_12\catcode `\%12\relax}%
\providecommand \@@startlink[1]{}%
\providecommand \@@endlink[0]{}%
\providecommand \url  [0]{\begingroup\@sanitize@url \@url }%
\providecommand \@url [1]{\endgroup\@href {#1}{\urlprefix }}%
\providecommand \urlprefix  [0]{URL }%
\providecommand \Eprint [0]{\href }%
\providecommand \doibase [0]{https://doi.org/}%
\providecommand \selectlanguage [0]{\@gobble}%
\providecommand \bibinfo  [0]{\@secondoftwo}%
\providecommand \bibfield  [0]{\@secondoftwo}%
\providecommand \translation [1]{[#1]}%
\providecommand \BibitemOpen [0]{}%
\providecommand \bibitemStop [0]{}%
\providecommand \bibitemNoStop [0]{.\EOS\space}%
\providecommand \EOS [0]{\spacefactor3000\relax}%
\providecommand \BibitemShut  [1]{\csname bibitem#1\endcsname}%
\let\auto@bib@innerbib\@empty
\bibitem [{\citenamefont {Feng}(2010)}]{Feng:2010gw}%
  \BibitemOpen
  \bibfield  {author} {\bibinfo {author} {\bibfnamefont {J.~L.}\ \bibnamefont {Feng}},\ }\bibfield  {title} {\bibinfo {title} {{Dark matter candidates from particle physics and methods of detection}},\ }\href {https://doi.org/10.1146/annurev-astro-082708-101659} {\bibfield  {journal} {\bibinfo  {journal} {Annu. Rev. Astron. Astrophys.}\ }\textbf {\bibinfo {volume} {48}},\ \bibinfo {pages} {495} (\bibinfo {year} {2010})}\BibitemShut {NoStop}%
\bibitem [{\citenamefont {Behnke}\ \emph {et~al.}(2012)\citenamefont {Behnke} \emph {et~al.}}]{COUPP:2012jrk}%
  \BibitemOpen
  \bibfield  {author} {\bibinfo {author} {\bibfnamefont {E.}~\bibnamefont {Behnke}} \emph {et~al.} (\bibinfo {collaboration} {COUPP Collaboration}),\ }\bibfield  {title} {\bibinfo {title} {{First dark matter search results from a 4-kg CF$_3$I bubble chamber operated in a deep underground site}},\ }\href {https://doi.org/10.1103/PhysRevD.86.052001} {\bibfield  {journal} {\bibinfo  {journal} {Phys. Rev. D}\ }\textbf {\bibinfo {volume} {86}},\ \bibinfo {pages} {052001} (\bibinfo {year} {2012})},\ \bibinfo {note} {; 90, 079902 (2014)}\BibitemShut {NoStop}%
\bibitem [{\citenamefont {Amole}\ \emph {et~al.}(2015)\citenamefont {Amole} \emph {et~al.}}]{PICO:2015yox}%
  \BibitemOpen
  \bibfield  {author} {\bibinfo {author} {\bibfnamefont {C.}~\bibnamefont {Amole}} \emph {et~al.} (\bibinfo {collaboration} {PICO Collaboration}),\ }\bibfield  {title} {\bibinfo {title} {{Dark matter search results from the PICO-2L C$_3$F$_8$ bubble chamber}},\ }\href {https://doi.org/10.1103/PhysRevLett.114.231302} {\bibfield  {journal} {\bibinfo  {journal} {Phys. Rev. Lett.}\ }\textbf {\bibinfo {volume} {114}},\ \bibinfo {pages} {231302} (\bibinfo {year} {2015})}\BibitemShut {NoStop}%
\bibitem [{\citenamefont {Behnke}\ \emph {et~al.}(2017)\citenamefont {Behnke} \emph {et~al.}}]{Behnke:2016lsk}%
  \BibitemOpen
  \bibfield  {author} {\bibinfo {author} {\bibfnamefont {E.}~\bibnamefont {Behnke}} \emph {et~al.},\ }\bibfield  {title} {\bibinfo {title} {{Final results of the PICASSO dark matter search experiment}},\ }\href {https://doi.org/10.1016/j.astropartphys.2017.02.005} {\bibfield  {journal} {\bibinfo  {journal} {Astropart. Phys.}\ }\textbf {\bibinfo {volume} {90}},\ \bibinfo {pages} {85} (\bibinfo {year} {2017})}\BibitemShut {NoStop}%
\bibitem [{\citenamefont {Amole}\ \emph {et~al.}(2017)\citenamefont {Amole} \emph {et~al.}}]{PICO:2017tgi}%
  \BibitemOpen
  \bibfield  {author} {\bibinfo {author} {\bibfnamefont {C.}~\bibnamefont {Amole}} \emph {et~al.} (\bibinfo {collaboration} {PICO Collaboration}),\ }\bibfield  {title} {\bibinfo {title} {{Dark matter search results from the PICO-60 C$_3$F$_8$ bubble chamber}},\ }\href {https://doi.org/10.1103/PhysRevLett.118.251301} {\bibfield  {journal} {\bibinfo  {journal} {Phys. Rev. Lett.}\ }\textbf {\bibinfo {volume} {118}},\ \bibinfo {pages} {251301} (\bibinfo {year} {2017})}\BibitemShut {NoStop}%
\bibitem [{\citenamefont {Amole}\ \emph {et~al.}(2019{\natexlab{a}})\citenamefont {Amole} \emph {et~al.}}]{PICO:2019vsc}%
  \BibitemOpen
  \bibfield  {author} {\bibinfo {author} {\bibfnamefont {C.}~\bibnamefont {Amole}} \emph {et~al.} (\bibinfo {collaboration} {PICO Collaboration}),\ }\bibfield  {title} {\bibinfo {title} {{Dark matter search results from the complete exposure of the PICO-60 C$_3$F$_8$ bubble chamber}},\ }\href {https://doi.org/10.1103/PhysRevD.100.022001} {\bibfield  {journal} {\bibinfo  {journal} {Phys. Rev. D}\ }\textbf {\bibinfo {volume} {100}},\ \bibinfo {pages} {022001} (\bibinfo {year} {2019}{\natexlab{a}})}\BibitemShut {NoStop}%
\bibitem [{\citenamefont {Baxter}\ \emph {et~al.}(2017)\citenamefont {Baxter} \emph {et~al.}}]{Baxter:2017ozv}%
  \BibitemOpen
  \bibfield  {author} {\bibinfo {author} {\bibfnamefont {D.}~\bibnamefont {Baxter}} \emph {et~al.},\ }\bibfield  {title} {\bibinfo {title} {{First demonstration of a scintillating xenon bubble chamber for detecting dark matter and coherent elastic neutrino-nucleus scattering}},\ }\href {https://doi.org/10.1103/PhysRevLett.118.231301} {\bibfield  {journal} {\bibinfo  {journal} {Phys. Rev. Lett.}\ }\textbf {\bibinfo {volume} {118}},\ \bibinfo {pages} {231301} (\bibinfo {year} {2017})}\BibitemShut {NoStop}%
\bibitem [{\citenamefont {Giampa}()}]{Giampa:2021wte}%
  \BibitemOpen
  \bibfield  {author} {\bibinfo {author} {\bibfnamefont {P.}~\bibnamefont {Giampa}} (\bibinfo {collaboration} {SBC Collaboration}),\ }\bibfield  {title} {\bibinfo {title} {{The scintillating bubble chamber (SBC) experiment for dark matter and reactor CEvNS}},\ }\href@noop {} {\bibinfo  {journal} {Pros. Sci. ICHEP2020 (2021) 632}\ }\BibitemShut {NoStop}%
\bibitem [{\citenamefont {Flores}\ \emph {et~al.}(2021)\citenamefont {Flores} \emph {et~al.}}]{SBC:2021yal}%
  \BibitemOpen
\bibfield  {journal} {  }\bibfield  {author} {\bibinfo {author} {\bibfnamefont {L.~J.}\ \bibnamefont {Flores}} \emph {et~al.} (\bibinfo {collaboration} {SBC Collaboration and CE\ensuremath{\nu}NS Theory Group at IF-UNAM}),\ }\bibfield  {title} {\bibinfo {title} {{Physics reach of a low threshold scintillating argon bubble chamber in coherent elastic neutrino-nucleus scattering reactor experiments}},\ }\href {https://doi.org/10.1103/PhysRevD.103.L091301} {\bibfield  {journal} {\bibinfo  {journal} {Phys. Rev. D}\ }\textbf {\bibinfo {volume} {103}},\ \bibinfo {pages} {L091301} (\bibinfo {year} {2021})}\BibitemShut {NoStop}%
\bibitem [{\citenamefont {Seitz}(1958)}]{Seitz:1958nva}%
  \BibitemOpen
  \bibfield  {author} {\bibinfo {author} {\bibfnamefont {F.}~\bibnamefont {Seitz}},\ }\bibfield  {title} {\bibinfo {title} {{On the theory of the bubble chamber}},\ }\href {https://doi.org/10.1063/1.1724333} {\bibfield  {journal} {\bibinfo  {journal} {Phys. Fluids}\ }\textbf {\bibinfo {volume} {1}},\ \bibinfo {pages} {2} (\bibinfo {year} {1958})}\BibitemShut {NoStop}%
\bibitem [{\citenamefont {Ali}\ \emph {et~al.}(2022)\citenamefont {Ali} \emph {et~al.}}]{PICO:2022nyi}%
  \BibitemOpen
  \bibfield  {author} {\bibinfo {author} {\bibfnamefont {B.}~\bibnamefont {Ali}} \emph {et~al.} (\bibinfo {collaboration} {PICO COllaboration}),\ }\bibfield  {title} {\bibinfo {title} {{Determining the bubble nucleation efficiency of low-energy nuclear recoils in superheated C3F8 dark matter detectors}},\ }\href {https://doi.org/10.1103/PhysRevD.106.122003} {\bibfield  {journal} {\bibinfo  {journal} {Phys. Rev. D}\ }\textbf {\bibinfo {volume} {106}},\ \bibinfo {pages} {122003} (\bibinfo {year} {2022})}\BibitemShut {NoStop}%
\bibitem [{\citenamefont {Pless}\ and\ \citenamefont {Plano}(1956)}]{Pless:1956}%
  \BibitemOpen
  \bibfield  {author} {\bibinfo {author} {\bibfnamefont {I.~A.}\ \bibnamefont {Pless}}\ and\ \bibinfo {author} {\bibfnamefont {R.~J.}\ \bibnamefont {Plano}},\ }\bibfield  {title} {\bibinfo {title} {{Negative pressure isopentane bubble chamber}},\ }\href {https://doi.org/10.1063/1.1715416} {\bibfield  {journal} {\bibinfo  {journal} {Rev. Sci. Instrum.}\ }\textbf {\bibinfo {volume} {27}},\ \bibinfo {pages} {935} (\bibinfo {year} {1956})}\BibitemShut {NoStop}%
\bibitem [{\citenamefont {Amole}\ \emph {et~al.}(2019{\natexlab{b}})\citenamefont {Amole} \emph {et~al.}}]{PICO:2019rsv}%
  \BibitemOpen
  \bibfield  {author} {\bibinfo {author} {\bibfnamefont {C.}~\bibnamefont {Amole}} \emph {et~al.} (\bibinfo {collaboration} {PICO Collaboration}),\ }\bibfield  {title} {\bibinfo {title} {{Data-driven modeling of electron recoil nucleation in PICO C$_3$F$_8$ bubble chambers}},\ }\href {https://doi.org/10.1103/PhysRevD.100.082006} {\bibfield  {journal} {\bibinfo  {journal} {Phys. Rev. D}\ }\textbf {\bibinfo {volume} {100}},\ \bibinfo {pages} {082006} (\bibinfo {year} {2019}{\natexlab{b}})}\BibitemShut {NoStop}%
\bibitem [{\citenamefont {Harper}(1993)}]{Harper_1993}%
  \BibitemOpen
  \bibfield  {author} {\bibinfo {author} {\bibfnamefont {M.~J.}\ \bibnamefont {Harper}},\ }\bibfield  {title} {\bibinfo {title} {{Calculation of recoil ion effective track lengths in neutron-radiation-induced nucleation}},\ }\href {https://doi.org/10.13182/NSE93-A24023} {\bibfield  {journal} {\bibinfo  {journal} {Nucl. Sci. and Eng.}\ }\textbf {\bibinfo {volume} {114}},\ \bibinfo {pages} {118} (\bibinfo {year} {1993})}\BibitemShut {NoStop}%
\bibitem [{\citenamefont {Denzel}\ \emph {et~al.}(2016)\citenamefont {Denzel}, \citenamefont {Diemand},\ and\ \citenamefont {Ang\'elil}}]{Denzel:2016epc}%
  \BibitemOpen
  \bibfield  {author} {\bibinfo {author} {\bibfnamefont {P.}~\bibnamefont {Denzel}}, \bibinfo {author} {\bibfnamefont {J.}~\bibnamefont {Diemand}},\ and\ \bibinfo {author} {\bibfnamefont {R.}~\bibnamefont {Ang\'elil}},\ }\bibfield  {title} {\bibinfo {title} {{Molecular dynamics simulations of bubble nucleation in dark matter detectors}},\ }\href {https://doi.org/10.1103/PhysRevE.93.013301} {\bibfield  {journal} {\bibinfo  {journal} {Phys. Rev. E}\ }\textbf {\bibinfo {volume} {93}},\ \bibinfo {pages} {013301} (\bibinfo {year} {2016})}\BibitemShut {NoStop}%
\bibitem [{\citenamefont {Frederic}(2019)}]{Tardif:2019}%
  \BibitemOpen
  \bibfield  {author} {\bibinfo {author} {\bibfnamefont {T.}~\bibnamefont {Frederic}},\ }\href@noop {} {\bibinfo {title} {{Direct detection of dark matter with the PICO experiment and the PICO-0.1 calibration chamber}}} (\bibinfo {year} {2019})\BibitemShut {NoStop}%
\bibitem [{\citenamefont {Plimpton}\ \emph {et~al.}(2021)\citenamefont {Plimpton}, \citenamefont {Kohlmeyer}, \citenamefont {Thompson}, \citenamefont {Moore},\ and\ \citenamefont {Berger}}]{Plimpton:2021}%
  \BibitemOpen
  \bibfield  {author} {\bibinfo {author} {\bibfnamefont {S.}~\bibnamefont {Plimpton}}, \bibinfo {author} {\bibfnamefont {A.}~\bibnamefont {Kohlmeyer}}, \bibinfo {author} {\bibfnamefont {A.}~\bibnamefont {Thompson}}, \bibinfo {author} {\bibfnamefont {S.}~\bibnamefont {Moore}},\ and\ \bibinfo {author} {\bibfnamefont {R.}~\bibnamefont {Berger}},\ }\href {https://doi.org/10.5281/zenodo.6386596} {\bibinfo {title} {{\lammp{} stable release 29 September 2021}}} (\bibinfo {year} {2021})\BibitemShut {NoStop}%
\bibitem [{\citenamefont {Kozynets}\ \emph {et~al.}(2019)\citenamefont {Kozynets}, \citenamefont {Fallows},\ and\ \citenamefont {Krauss}}]{Kozynets:2019ihv}%
  \BibitemOpen
  \bibfield  {author} {\bibinfo {author} {\bibfnamefont {T.}~\bibnamefont {Kozynets}}, \bibinfo {author} {\bibfnamefont {S.}~\bibnamefont {Fallows}},\ and\ \bibinfo {author} {\bibfnamefont {C.~B.}\ \bibnamefont {Krauss}},\ }\bibfield  {title} {\bibinfo {title} {{Modeling emission of acoustic energy during bubble expansion in PICO bubble chambers}},\ }\href {https://doi.org/10.1103/PhysRevD.100.052001} {\bibfield  {journal} {\bibinfo  {journal} {Phys. Rev. D}\ }\textbf {\bibinfo {volume} {100}},\ \bibinfo {pages} {052001} (\bibinfo {year} {2019})}\BibitemShut {NoStop}%
\bibitem [{\citenamefont {Errington}\ \emph {et~al.}(2003)\citenamefont {Errington}, \citenamefont {Debenedetti},\ and\ \citenamefont {Torquato}}]{Errington:2003}%
  \BibitemOpen
  \bibfield  {author} {\bibinfo {author} {\bibfnamefont {J.~R.}\ \bibnamefont {Errington}}, \bibinfo {author} {\bibfnamefont {P.~G.}\ \bibnamefont {Debenedetti}},\ and\ \bibinfo {author} {\bibfnamefont {S.}~\bibnamefont {Torquato}},\ }\bibfield  {title} {\bibinfo {title} {{Quantification of order in the Lennard-Jones system}},\ }\href {https://doi.org/10.1063/1.1532344} {\bibfield  {journal} {\bibinfo  {journal} {J. Chem. Phys.}\ }\textbf {\bibinfo {volume} {118}},\ \bibinfo {pages} {2256} (\bibinfo {year} {2003})}\BibitemShut {NoStop}%
\bibitem [{\citenamefont {Mikic}\ \emph {et~al.}(1970)\citenamefont {Mikic}, \citenamefont {Rohsenow},\ and\ \citenamefont {Griffith}}]{MIKIC1970657}%
  \BibitemOpen
  \bibfield  {author} {\bibinfo {author} {\bibfnamefont {B.}~\bibnamefont {Mikic}}, \bibinfo {author} {\bibfnamefont {W.}~\bibnamefont {Rohsenow}},\ and\ \bibinfo {author} {\bibfnamefont {P.}~\bibnamefont {Griffith}},\ }\bibfield  {title} {\bibinfo {title} {{On bubble growth rates}},\ }\href {https://doi.org/https://doi.org/10.1016/0017-9310(70)90040-2} {\bibfield  {journal} {\bibinfo  {journal} {Int. J. Heat and Mass Transfer}\ }\textbf {\bibinfo {volume} {13}},\ \bibinfo {pages} {657} (\bibinfo {year} {1970})}\BibitemShut {NoStop}%
\bibitem [{\citenamefont {Rayleigh}(1917)}]{rayleigh_1917}%
  \BibitemOpen
  \bibfield  {author} {\bibinfo {author} {\bibfnamefont {L.}~\bibnamefont {Rayleigh}},\ }\bibfield  {title} {\bibinfo {title} {{VIII. On the pressure developed in a liquid during the collapse of a spherical cavity }},\ }\href {https://doi.org/10.1080/14786440808635681} {\bibfield  {journal} {\bibinfo  {journal} {London, Edinburgh, Dublin Philos. Mag. J. Sci.}\ }\textbf {\bibinfo {volume} {34}},\ \bibinfo {pages} {94} (\bibinfo {year} {1917})}\BibitemShut {NoStop}%
\bibitem [{\citenamefont {Alfonso-Pita}\ \emph {et~al.}(2023)\citenamefont {Alfonso-Pita} \emph {et~al.}}]{Alfonso-Pita:2023frp}%
  \BibitemOpen
  \bibfield  {author} {\bibinfo {author} {\bibfnamefont {E.}~\bibnamefont {Alfonso-Pita}} \emph {et~al.},\ }\bibfield  {title} {\bibinfo {title} {{Scintillating bubble chambers for rare event searches}},\ }\href {https://doi.org/10.3390/universe9080346} {\bibfield  {journal} {\bibinfo  {journal} {Universe}\ }\textbf {\bibinfo {volume} {9}},\ \bibinfo {pages} {346} (\bibinfo {year} {2023})}\BibitemShut {NoStop}%
\bibitem [{\citenamefont {Amole}\ \emph {et~al.}(2016)\citenamefont {Amole} \emph {et~al.}}]{Amole_2016}%
  \BibitemOpen
  \bibfield  {author} {\bibinfo {author} {\bibfnamefont {C.}~\bibnamefont {Amole}} \emph {et~al.},\ }\bibfield  {title} {\bibinfo {title} {{Dark matter search results from the {PICO}-60 CF$_3$I bubble chamber}},\ }\href@noop {} {\bibfield  {journal} {\bibinfo  {journal} {Phys. Rev. D}\ }\textbf {\bibinfo {volume} {93}} (\bibinfo {year} {2016})}\BibitemShut {NoStop}%
\bibitem [{\citenamefont {Durnford}\ and\ \citenamefont {Piro}()}]{Durnford:2021cvb}%
  \BibitemOpen
  \bibfield  {author} {\bibinfo {author} {\bibfnamefont {D.}~\bibnamefont {Durnford}}\ and\ \bibinfo {author} {\bibfnamefont {M.-C.}\ \bibnamefont {Piro}} (\bibinfo {collaboration} {SBC and PICO Collaborations}),\ }\bibfield  {title} {\bibinfo {title} {{Nucleation efficiency of nuclear recoils in bubble chambers}},\ }\href@noop {} {\bibfield  {journal} {\bibinfo  {journal} {J. Instrum.}\ }\textbf {\bibinfo {volume} {17}}\bibinfo  {number} { (01)},\ \bibinfo {pages} {C01030 (2021)}}\BibitemShut {NoStop}%
\bibitem [{\citenamefont {Alfonso-Pita}\ \emph {et~al.}()\citenamefont {Alfonso-Pita} \emph {et~al.}}]{alfonsopita2022snowmass}%
  \BibitemOpen
\bibfield  {number} {  }\bibfield  {author} {\bibinfo {author} {\bibfnamefont {E.}~\bibnamefont {Alfonso-Pita}} \emph {et~al.},\ }\href@noop {} {\bibinfo {title} {{Snowmass 2021 scintillating bubble chambers: Liquid-noble bubble chambers for dark matter and CE$\nu$NS detection}}},\ \Eprint {https://arxiv.org/abs/2207.12400} {arXiv:2207.12400} \BibitemShut {NoStop}%
\bibitem [{\citenamefont {Linstrom}\ and\ \citenamefont {W.G.~Mallard}()}]{NIST:FIPS1402}%
  \BibitemOpen
  \bibfield  {author} {\bibinfo {author} {\bibfnamefont {P.}~\bibnamefont {Linstrom}}\ and\ \bibinfo {author} {\bibfnamefont {E.}~\bibnamefont {W.G.~Mallard}},\ }\href@noop {} {\bibinfo {title} {{{in NIST Chemistry WebBook, NIST Standard Reference Database Number 69}}}},\ \bibinfo {note} {{(NIST, Gaithersburg MD), Vol. 69, \url{http://webbook.nist.gov} (access January 2024)}}\BibitemShut {NoStop}%
\bibitem [{\citenamefont {Lemmon}\ \emph {et~al.}(2018)\citenamefont {Lemmon}, \citenamefont {Bell}, \citenamefont {Huber},\ and\ \citenamefont {McLinden}}]{LEMMON-RP10}%
  \BibitemOpen
  \bibfield  {author} {\bibinfo {author} {\bibfnamefont {E.~W.}\ \bibnamefont {Lemmon}}, \bibinfo {author} {\bibfnamefont {I.~H.}\ \bibnamefont {Bell}}, \bibinfo {author} {\bibfnamefont {M.~L.}\ \bibnamefont {Huber}},\ and\ \bibinfo {author} {\bibfnamefont {M.~O.}\ \bibnamefont {McLinden}},\ }\href {https://doi.org/https://doi.org/10.18434/T4/1502528} {\bibinfo {title} {{{NIST Standard Reference Database 23: Reference Fluid Thermodynamic and Transport Properties--\refprop{}, Version 10.0}}}} (\bibinfo {year} {NIST, Gaithersburg, 2018})\BibitemShut {NoStop}%
\bibitem [{\citenamefont {Ziegler}\ \emph {et~al.}(2010)\citenamefont {Ziegler}, \citenamefont {Ziegler},\ and\ \citenamefont {Biersack}}]{Ziegler:2010}%
  \BibitemOpen
  \bibfield  {author} {\bibinfo {author} {\bibfnamefont {J.~F.}\ \bibnamefont {Ziegler}}, \bibinfo {author} {\bibfnamefont {M.}~\bibnamefont {Ziegler}},\ and\ \bibinfo {author} {\bibfnamefont {J.}~\bibnamefont {Biersack}},\ }\bibfield  {title} {\bibinfo {title} {{SRIM – the stopping and range of ions in matter (2010)}},\ }\href {https://doi.org/10.1016/j.nimb.2010.02.091} {\bibfield  {journal} {\bibinfo  {journal} {Nucl. Instrum. and Methods Phys. Res. Sect. B}\ }\textbf {\bibinfo {volume} {268}},\ \bibinfo {pages} {1818} (\bibinfo {year} {2010})}\BibitemShut {NoStop}%
\bibitem [{\citenamefont {Lindhard}\ and\ \citenamefont {Scharff}(1961)}]{Lindhard:1961zz}%
  \BibitemOpen
  \bibfield  {author} {\bibinfo {author} {\bibfnamefont {J.}~\bibnamefont {Lindhard}}\ and\ \bibinfo {author} {\bibfnamefont {M.}~\bibnamefont {Scharff}},\ }\bibfield  {title} {\bibinfo {title} {{Energy dissipation by ions in the kev region}},\ }\href {https://doi.org/10.1103/PhysRev.124.128} {\bibfield  {journal} {\bibinfo  {journal} {Phys. Rev.}\ }\textbf {\bibinfo {volume} {124}},\ \bibinfo {pages} {128} (\bibinfo {year} {1961})}\BibitemShut {NoStop}%
\bibitem [{\citenamefont {Hitachi}(2008)}]{Hitachi:2008kf}%
  \BibitemOpen
  \bibfield  {author} {\bibinfo {author} {\bibfnamefont {A.}~\bibnamefont {Hitachi}},\ }\bibfield  {title} {\bibinfo {title} {{Bragg-like curve for dark matter searches: binary gases}},\ }\href {https://doi.org/10.1016/j.radphyschem.2008.05.044} {\bibfield  {journal} {\bibinfo  {journal} {Radiat. Phys. Chem.}\ }\textbf {\bibinfo {volume} {77}},\ \bibinfo {pages} {1311} (\bibinfo {year} {2008})}\BibitemShut {NoStop}%
\bibitem [{\citenamefont {Szydagis}\ \emph {et~al.}(2023)\citenamefont {Szydagis} \emph {et~al.}}]{NEST:2023}%
  \BibitemOpen
  \bibfield  {author} {\bibinfo {author} {\bibfnamefont {M.}~\bibnamefont {Szydagis}} \emph {et~al.},\ }\href {https://doi.org/10.5281/zenodo.8215927} {\bibinfo {title} {{Noble element simulation technique}}} (\bibinfo {year} {2023})\BibitemShut {NoStop}%
\bibitem [{\citenamefont {Behnke}\ \emph {et~al.}(2013)\citenamefont {Behnke} \emph {et~al.}}]{COUPP:2013yjn}%
  \BibitemOpen
  \bibfield  {author} {\bibinfo {author} {\bibfnamefont {E.}~\bibnamefont {Behnke}} \emph {et~al.} (\bibinfo {collaboration} {COUPP Collaboration}),\ }\bibfield  {title} {\bibinfo {title} {{Direct measurement of the bubble nucleation energy threshold in a CF$_3$I bubble chamber}},\ }\href {https://doi.org/10.1103/PhysRevD.88.021101} {\bibfield  {journal} {\bibinfo  {journal} {Phys. Rev. D}\ }\textbf {\bibinfo {volume} {88}},\ \bibinfo {pages} {021101} (\bibinfo {year} {2013})}\BibitemShut {NoStop}%
\bibitem [{\citenamefont {Jackson}\ \emph {et~al.}(2023)\citenamefont {Jackson}, \citenamefont {Savoie}, \citenamefont {Statt},\ and\ \citenamefont {Webb}}]{ML_MD}%
  \BibitemOpen
  \bibfield  {author} {\bibinfo {author} {\bibfnamefont {N.~E.}\ \bibnamefont {Jackson}}, \bibinfo {author} {\bibfnamefont {B.~M.}\ \bibnamefont {Savoie}}, \bibinfo {author} {\bibfnamefont {A.}~\bibnamefont {Statt}},\ and\ \bibinfo {author} {\bibfnamefont {M.~A.}\ \bibnamefont {Webb}},\ }\bibfield  {title} {\bibinfo {title} {Introduction to machine learning for molecular simulation},\ }\href {https://doi.org/10.1021/acs.jctc.3c00735} {\bibfield  {journal} {\bibinfo  {journal} {J. Chem. Theor. Comput.}\ }\textbf {\bibinfo {volume} {19}},\ \bibinfo {pages} {4335} (\bibinfo {year} {2023})}\BibitemShut {NoStop}%
\bibitem [{\citenamefont {Zhang}\ \emph {et~al.}(2023)\citenamefont {Zhang} \emph {et~al.}}]{AI_MD}%
  \BibitemOpen
  \bibfield  {author} {\bibinfo {author} {\bibfnamefont {J.}~\bibnamefont {Zhang}} \emph {et~al.},\ }\bibfield  {title} {\bibinfo {title} {Artificial intelligence enhanced molecular simulations},\ }\href {https://doi.org/10.1021/acs.jctc.3c00214} {\bibfield  {journal} {\bibinfo  {journal} {J. Chem. Theor. Comput.}\ }\textbf {\bibinfo {volume} {19}},\ \bibinfo {pages} {4338} (\bibinfo {year} {2023})}\BibitemShut {NoStop}%
\bibitem [{\citenamefont {Feldman}\ and\ \citenamefont {Cousins}(1998)}]{Feldman_1998}%
  \BibitemOpen
  \bibfield  {author} {\bibinfo {author} {\bibfnamefont {G.~J.}\ \bibnamefont {Feldman}}\ and\ \bibinfo {author} {\bibfnamefont {R.~D.}\ \bibnamefont {Cousins}},\ }\bibfield  {title} {\bibinfo {title} {{Unified approach to the classical statistical analysis of small signals}},\ }\href {https://doi.org/10.1103/physrevd.57.3873} {\bibfield  {journal} {\bibinfo  {journal} {Physical Review D}\ }\textbf {\bibinfo {volume} {57}},\ \bibinfo {pages} {3873} (\bibinfo {year} {1998})}\BibitemShut {NoStop}%
\bibitem [{\citenamefont {Gluscevic}\ \emph {et~al.}(2015)\citenamefont {Gluscevic}, \citenamefont {Gresham}, \citenamefont {McDermott}, \citenamefont {Peter},\ and\ \citenamefont {Zurek}}]{dmdd_Gluscevic_2015}%
  \BibitemOpen
  \bibfield  {author} {\bibinfo {author} {\bibfnamefont {V.}~\bibnamefont {Gluscevic}}, \bibinfo {author} {\bibfnamefont {M.~I.}\ \bibnamefont {Gresham}}, \bibinfo {author} {\bibfnamefont {S.~D.}\ \bibnamefont {McDermott}}, \bibinfo {author} {\bibfnamefont {A.~H.}\ \bibnamefont {Peter}},\ and\ \bibinfo {author} {\bibfnamefont {K.~M.}\ \bibnamefont {Zurek}},\ }\bibfield  {title} {\bibinfo {title} {{Identifying the theory of dark matter with direct detection}},\ }\href {https://doi.org/10.1088/1475-7516/2015/12/057} {\bibfield  {journal} {\bibinfo  {journal} {J. Cosmol. Astropart. Phys. 12}\ ,\ \bibinfo {pages} {057}} (\bibinfo {year} {2015})}\BibitemShut {NoStop}%
\bibitem [{\citenamefont {Lewin}\ and\ \citenamefont {Smith}(1996)}]{Lewin:1995rx}%
  \BibitemOpen
  \bibfield  {author} {\bibinfo {author} {\bibfnamefont {J.~D.}\ \bibnamefont {Lewin}}\ and\ \bibinfo {author} {\bibfnamefont {P.~F.}\ \bibnamefont {Smith}},\ }\bibfield  {title} {\bibinfo {title} {{Review of mathematics, numerical factors, and corrections for dark matter experiments based on elastic nuclear recoil}},\ }\href {https://doi.org/10.1016/S0927-6505(96)00047-3} {\bibfield  {journal} {\bibinfo  {journal} {Astropart. Phys.}\ }\textbf {\bibinfo {volume} {6}},\ \bibinfo {pages} {87} (\bibinfo {year} {1996})}\BibitemShut {NoStop}%
\bibitem [{ont()}]{ontario}%
  \BibitemOpen
  \href@noop {} {}\bibinfo {howpublished} {See \url{https://computeontario.ca}}\BibitemShut {NoStop}%
\bibitem [{all()}]{alliance}%
  \BibitemOpen
  \href@noop {} {}\bibinfo {howpublished} {See \url{https://www.alliancecan.ca}}\BibitemShut {NoStop}%
\bibitem [{\citenamefont {Lindhard}\ \emph {et~al.}(1963)\citenamefont {Lindhard}, \citenamefont {Nielsen}, \citenamefont {Scharff},\ and\ \citenamefont {Thomsen}}]{Lindhard:1963}%
  \BibitemOpen
  \bibfield  {author} {\bibinfo {author} {\bibfnamefont {J.}~\bibnamefont {Lindhard}}, \bibinfo {author} {\bibfnamefont {V.}~\bibnamefont {Nielsen}}, \bibinfo {author} {\bibfnamefont {M.}~\bibnamefont {Scharff}},\ and\ \bibinfo {author} {\bibfnamefont {P.~V.}\ \bibnamefont {Thomsen}},\ }\bibfield  {title} {\bibinfo {title} {{Integral equations governing radiation effects. (Notes on atomic collisions, III)}},\ }\href {https://www.osti.gov/biblio/4701226} {\bibfield  {journal} {\bibinfo  {journal} {K. Dan. Vidensk. Selsk., Mat.-Fys. Medd.}\ }\textbf {\bibinfo {volume} {33}} (\bibinfo {year} {1963})}\BibitemShut {NoStop}%
\bibitem [{\citenamefont {Szydagis}\ \emph {et~al.}(2021)\citenamefont {Szydagis} \emph {et~al.}}]{Szydagis:2021hfh}%
  \BibitemOpen
  \bibfield  {author} {\bibinfo {author} {\bibfnamefont {M.}~\bibnamefont {Szydagis}} \emph {et~al.},\ }\bibfield  {title} {\bibinfo {title} {{A review of basic energy reconstruction techniques in liquid xenon and argon detectors for dark matter and neutrino physics using \nest{}}},\ }\href {https://doi.org/10.3390/instruments5010013} {\bibfield  {journal} {\bibinfo  {journal} {Instruments}\ }\textbf {\bibinfo {volume} {5}},\ \bibinfo {pages} {13} (\bibinfo {year} {2021})}\BibitemShut {NoStop}%
\end{thebibliography}%

\end{document}